# Complete electronic phase diagram and enhanced superconductivity in fluorine-doped PrFeAsO$_{1-x}$F$_x$

Priya Singh[1], Konrad Kwatek[2], Tatiana Zajarniuk[3], Taras Palasyuk[2], Cezariusz Jastrzębski[2], A. Szewczyk[3], Shiv J. Singh[1]*

[1]*Institute of High Pressure Physics (IHPP), Polish Academy of Sciences, Sokołowska 29/37, 01-142 Warsaw, Poland*

[2]*Faculty of Physics, Warsaw University of Technology, Koszykowa 75, 00-662, Warsaw, Poland*

[3]*Institute of Physics, Polish Academy of Sciences, Aleja Lotnikow 32/46, 02-668 Warsaw, Poland*

**\*Corresponding author:**
 Email: sjs@unipress.waw.pl
 https://orcid.org/0000-0001-5769-1787

# Abstract

Establishing a complete electronic phase diagram for *RE*FeAsO (*RE* = rare earth, *RE*1111) iron-based superconductors has remained experimentally challenging. Here, we report a systematic investigation of PrFeAsO$_{1-x}$F$_x$ over the full nominal fluorine-doping range $0 \leq x \leq 1$ and construct the first comprehensive electronic phase diagram for this system. The evolution from the nonsuperconducting parent compound to the fluorine-rich limit reveals a broad dome-shaped superconducting region. Structural refinement demonstrates a systematic lattice contraction with increasing fluorine content (*x*), corroborated by Raman spectroscopy through softening of the Pr(A$_{1g}$) phonon mode and hardening of the Fe(B$_{1g}$) mode, confirming effective fluorine incorporation at the oxygen sites. The maximum superconducting transition temperature ($T_c$) reaches up to 52.3 K, approximately 5 K higher than previous reports for Pr1111. Magnetotransport measurements yield large upper critical fields $H_{c2}(0)$ exceeding 100 T, while analysis of resistive transition broadening reveals thermally activated flux flow with a crossover from single-vortex to collective pinning regimes. Specific-heat measurements exhibit a reduced jump $\Delta C/\gamma T_c < 1.43$, reflecting strong superconducting fluctuations and multiband pairing. These results establish clear structure–property correlations and provide a unified description of superconductivity across the entire doping range of the Pr1111 system.

***Keywords:*** Iron based superconductors, Critical transition temperature, Critical current density, Superconducting phase diagram

# I. INTRODUCTION

The discovery of superconductivity in LaFeAsO$_{1-x}$F$_x$ with a transition temperature $T_c$ = 26 K by Kamihara *et al.* in 2008 [1] marked the advent of iron-based superconductors (IBS) and initiated extensive research into their unconventional pairing mechanisms. This breakthrough led to the identification of the *RE*FeAsO (*RE* = rare earth) compounds belonging to IBS—commonly referred to as the 1111 family (*RE*1111)—which includes La, Ce, Pr, Nd, Sm, and Gd members [2]. In their undoped state, *RE*1111 compounds undergo coupled structural and spin-density-wave (SDW) transitions near 130–140 K and are nonsuperconducting. Superconductivity can be induced by suppressing these instabilities via carrier doping, achieved through electron doping at the oxygen site (F or H substitution) [3], transition-metal substitution at the Fe site [4], [5], isovalent substitution at the As site [6], [7], or hole doping at the rare-earth site [8]. In most cases, superconductivity emerges over a limited doping range, with the maximum $T_c$ typically occurring near ∼20% fluorine substitution, as observed in SmFeAsO$_{1-x}$F$_x$ [9]. Extending the doping range beyond this level, however, remains challenging due to impurity phase formation and fluorine volatility during conventional solid-state synthesis process, which degrade phase purity and superconducting properties [10].

Among IBS, the 1111 family exhibits the highest superconducting transition temperatures, reaching $T_c \approx$ 58 K in fluorine-doped SmFeAsO [9], together with exceptionally large upper critical fields ($H_{c2} \geq$ 100 T) and high critical current densities ($J_c$~$10^6$A/cm$^2$), making these materials promising for high-field applications. Despite extensive studies on La1111, Sm1111, Nd1111, and Ce1111, comparatively few investigations have focused on PrFeAsO largely due to its demanding synthesis conditions. In most reports, fluorine doping in 1111 compounds (*RE*FeAsO$_{1-x}$F$_x$) is limited to $x \lesssim$ 0.2–0.3, resulting in incomplete electronic phase diagrams [11]. High-pressure synthesis technique [10] has enabled hydrogen substitution up to $x \approx$ 0.5–0.6 in *RE*FeAsO$_{1-x}$H$_x$, revealing complex phase diagrams with one or two superconducting domes [12]. However, the role of hydrogen remains controversial, as superconductivity may also be influenced by oxygen deficiency, complicating the interpretation of the intrinsic electronic phase diagram. A complete phase diagram based on other dopants such as fluorine substitution—free from such ambiguities—is therefore highly desirable.

To further explore the 1111 family, we have selected F-doped PrFeAsO (Pr1111) as the subject of the present study. PrFeAsO occupies a particularly intriguing position within the 1111 family due to the interplay between Fe and rare-earth magnetism. Similar to other *RE*1111 compounds, PrFeAsO undergoes a tetragonal-to-orthorhombic structural transition at $T_s \sim 140$ K, followed by antiferromagnetic ordering of Fe moments at $T_N^{Fe}$. In addition, the $Pr^{3+}$ moments order antiferromagnetically at $T_N^{Pr} \approx 12$ K, accompanied by a reorientation of Fe spins along the crystallographic *c*-axis [13], [14], [15]. This low-temperature magnetic ordering introduces additional scattering channels in the FeAs layers and provides a unique platform to study the coupling between rare-earth magnetism and superconductivity. Previous studies employing angle-resolved photoemission spectroscopy [16], transport [17], neutron diffraction [18], [19], [20], specific heat [13], [21], and pressure-dependent measurements [22] have explored selected compositions of PrFeAsO$_{1-x}$F$_x$, but these investigations were restricted to narrow doping ranges ($x \approx 0.1$–0.2). Although superconductivity was reported as early as 2009 by Rotundu *et al.* [23], with a maximum $T_c \sim 47$ K near $x = 0.15$-0.35, a comprehensive electronic and superconducting phase diagram for Pr1111 has remained unresolved.

In this work, we address this long-standing gap by synthesizing a complete series of polycrystalline PrFeAsO$_{1-x}$F$_x$ samples spanning the full fluorine concentration range ($0 \leq x \leq 1$) using a conventional solid-state reaction method. A systematic investigation combining structural, transport, magnetic, Raman, and thermodynamic measurements under magnetic fields up to 9 T has been carried out. Superconductivity emerges at $x = 0.15$, coincident with the suppression of the structural and SDW transitions, and persists up to $x \approx 0.7$, where a maximum $T_c \approx 52.3$ K is achieved. Magnetotransport measurements reveal extremely large upper critical fields, $H_{c2}(0) \approx 250$ T for PrFeAsO$_{0.65}$F$_{0.35}$ and $\approx 212$ T for PrFeAsO$_{0.7}$F$_{0.3}$, corresponding to short coherence lengths and significant Pauli-limiting effects. Based on these results, we construct the first complete superconducting phase diagram of PrFeAsO$_{1-x}$F$_x$, delineating the underdoped ($0.15 \leq x < 0.3$), optimal doped ($0.3 \leq x \leq 0.4$), and overdoped ($0.4 < x \leq 0.7$) regimes. These findings provide new insight into the doping evolution, magnetic interactions, and superconducting mechanism in the Pr1111 system and advance the broader understanding of high-$T_c$ superconductivity in the 1111 family.

## II. EXPERIMENTAL METHODS

Polycrystalline PrFeAsO$_{1-x}$F$_x$ samples with the nominal fluorine contents $x = 0, 0.1, 0.15, 0.2, 0.25, 0.3, 0.35, 0.4, 0.5, 0.6, 0.7, 0.8$ and $1.0$ were synthesized using a two-step conventional solid-state reaction method. High purity Pr (3N), PrF$_3$ (3N), Fe$_2$O$_3$ (4N), As chunks (6N), and Fe (4N) powders were weighed according to nominal stoichiometry and thoroughly mixed for 10-15 mins using an agate mortar and pestle inside a high-purity argon-filled glove box to avoid oxidation and moisture contamination. The resulting homogeneous mixtures were cold-pressed into 11 mm-diameter pellets under a pressure of approximately 200 bars, sealed in tantalum tubes, and subsequently encapsulated in evacuated quartz tubes ($10^{-2}$-$10^{-3}$ bar). The samples were first heat-treated at 550 °C for 8 h, then reground, repressed, resealed under identical vacuum conditions, and annealed at 950 °C for 12 h, followed by controlled cooling to 110 °C over 18 h and furnace cooling to room temperature. The resulting dense polycrystalline pellets were mechanically cut into rectangular bars for subsequent structural, transport, magnetic, Raman, and specific heat characterizations. The synthesis was carried out in two steps, as described above. While additional grinding and annealing cycles can improve grain connectivity and densification, excessive thermal processing may cause partial fluorine loss due to volatilization at elevated temperatures, thereby reducing the effective doping level and suppressing superconductivity, as reported for other F-doped 1111 compounds [10], [24]. To mitigate this effect, our synthesis was limited to two-steps process.

Structural analysis has been performed by powder X-ray diffraction (XRD) measurements, which were carried out by using a Panalytical Empyrean diffractometer within the $2\theta$ range of 10° to 90°, with a step size of 0.013° and a counting time of 300 seconds per step, utilizing CuK$\alpha$ radiation ($\lambda$ = 1.5418 Å) at 35 mA and 40 kV. The analysis of the measured data was carried out using PDF4+2025 database through the ICDD database to extract the profile and assess the phase purity of the measured data. Magnetic measurements were performed by a vibrating sample magnetometer attached to a physical property measurement system (PPMS, Quantum Design). The temperature dependence of magnetic susceptibility was measured in zero-field-cooled (ZFC) and field-cooled (FC) modes under an applied magnetic field of 20 Oe in the temperature range of 5 to 60 K. The magnetic hysteresis loops at 5 K were recorded under the applied magnetic fields up to 9 T. Electrical resistivity was measured using a standard four-probe technique on rectangular specimens with typical dimensions of approximately 3 × 3 × 2 mm³. Detailed sample dimension is provided in the Supplementary Table T1. Electrical contacts were made using silver paste and copper wires. The zero-field resistivity measurements were carried out between 7-300 K using either a closed-cycle refrigerator or the PPMS. Magnetotransport and specific heat measurements for selected compositions were performed in the PPMS at temperatures below 200 K under magnetic fields up to 9 T. Raman scattering measurements were carried out using a LabRam ARAMIS (Horiba Jobin Yvon) spectrometer equipped with 632.8 nm wavelength of a He-Ne ion laser as the excitation source, further experimental details are provided elsewhere [25]. The incident laser beam was focused onto the sample surface through a 100× objective lens (numerical aperture, NA = 0.95), producing a spot size smaller than 3 μm. Backscattered light was collected by the same objective, dispersed using a 2400 lines/mm grating, and detected with a CCD over the Stokes shift range of 95-300 cm$^{-1}$. The laser power was limited to ~140 μW to minimize local heating, and spectra were acquired with an accumulation time of 300 s. Measurements were conducted directly on as-synthesized surfaces. Owing to the polycrystalline nature of the samples, spectra were collected from at least five different locations per composition, and Lorentzian fittings were employed to extract phonon peak positions associated with the PrFeAsO$_{1-x}$F$_x$ phase.

## III. RESULTS AND DISCUSSIONS
### a) Structural Analysis

The powder XRD patterns of all synthesized PrFeAsO$_{1-x}$F$_x$ samples are shown in Figure 1, separated into low- (0 ≤ $x$ ≤ 0.4) and high- (0.5 ≤ $x$ ≤ 1) doping regimes in Figures 1(a) and 1(b), respectively. All samples with 0 ≤ $x$ ≤ 0.7 crystallize predominantly in the tetragonal PrFeAsO-type structure (space group *P4/nmm*). The undoped compound and lightly doped samples with $x$ = 0.1 and 0.15 exhibit good phase purity. For higher fluorine contents, secondary phases including PrAs, PrOF, and FeAs gradually emerge, with their fraction increasing progressively at higher nominal $x$. The increase in fluorine content beyond $x$ = 0.5 leads to a pronounced growth of secondary phases, particularly PrOF and PrAs. For nominal fluorine contents $x$ ≥ 0.8, the tetragonal phase is strongly suppressed, and the samples are dominated by secondary phases, including PrAs, PrOF, PrF$_3$, and Fe$_2$As impurities. The observed diffraction patterns are analyzed by Rietveld refinement using the GSAS software for all compositions and a representative refinement for $x$ = 0.3 sample is shown in Figure 1(c). The refined lattice parameters '*a*' and '*c*', and the unit-cell volume '*V*' with the nominal fluorine content ($x$) is plotted in Figures 1(d)-(f) for all samples where the tetragonal PrFeAsO$_{1-x}$F$_x$ phase dominates. For comparison, we have also included the data from the previous reports [26] and [18] corresponding to $x$ = 0, 0.1, 0.15 and $x$ = 0.2 in these figures. The parent sample $x$ = 0 exhibits the lattice parameter $a = b$ = 3.978(68) Å and $c$ = 8.607(193) Å, consistent with the previous reports [18], [19], [26]. Both the lattice parameters (*a* and *c*) decrease systematically with increasing fluorine content up to $x$ ~ 0.5, reflecting the successful incorporation of smaller ionic radius of F$^-$ (~1.31 Å) compared to O$^{-2}$ (~1.38 Å) at oxygen sites.

For higher nominal fluorine concentrations $x > 0.5$, the lattice parameters and unit-cell volume remain nearly constant, suggesting saturation of fluorine incorporation into the superconducting tetragonal PrFeAsO$_{1-x}$F$_x$ phase, with excess fluorine accommodated in secondary phases such as PrOF. For the compositions $x = 0.8$ and $1.0$, the samples crystallize predominantly in a cubic phase with the space group *Fm-3m* and lattice parameters $a = b = c = 6.032(7)$ Å. As these compositions represent distinctly different crystal structures, they are not included in Figures 1(d)–(f). The overall results are consistent with previous studies on other members of the 1111 family [18], [19], [26], exhibiting a systematic lattice contraction and a corresponding reduction in unit cell volume with increasing fluorine content. These findings identify an effective fluorine solubility limit of approximately $x \approx 0.7$ in the PrFeAsO$_{1-x}$F$_x$ system, significantly exceeding the fluorine doping levels ($x \leq 0.3$) typically reported for F-doped *RE*1111 compounds [3]. Furthermore, the nature and evolution of impurity phases observed here are consistent with prior reports on F-doped *RE*1111 systems [10]. To roughly estimate the actual fluorine content in our PrFeAsO$_{1-x}$F$_x$ samples, we have compared the nominal fluorine concentrations ($x$) with those reported for Pr1111 [23], in which the actual fluorine substitution relative to the nominal value up to $x = 0.35$ was quantitatively determined using wavelength-dispersive spectroscopy (WDS). Based on this comparison, the nominal fluorine contents of $x = 0.15$, $0.20$, $0.25$, $0.30$, and $0.35$ in the present study correspond approximately to actual fluorine concentrations of $x = 0.11$, $0.15$, $0.19$, $0.20$, and $0.23$, respectively. Although these values should be regarded as approximate, they confirm substantial fluorine incorporation and support the extended solubility range achieved in the present synthesis.

b) **Raman spectroscopy**

The room-temperature unpolarized Raman spectra of pristine ($x = 0$) and fluorine-doped PrFeAsO$_{1-x}$F$_x$ are presented in Figure 2(a). The pristine PrFeAsO compound exhibits well-defined phonon modes corresponding to out-of-plane lattice vibrations of Pr(A$_{1g}$), As(A$_{1g}$), and Fe(B$_{1g}$) atoms. To the best of our knowledge, Raman spectra of PrFeAsO have not been previously reported; therefore, the phonon mode assignments are made by comparison with those of closely related *RE*FeAsO compounds [27], [28]. Fluorine substitution up to $x = 0.70$ does not produce qualitative changes in the Raman spectra, indicating that the tetragonal crystal structure and the overall bonding framework remain intact throughout the superconducting doping range. Nevertheless, systematic and composition-dependent shifts in phonon frequencies are observed, as summarized in Figure 2(b). With increasing fluorine content, the Pr($A_{1g}$) mode exhibits a pronounced softening of approximately 10 cm⁻¹, whereas the Fe($B_{1g}$) mode shows a modest hardening of about 4 cm⁻¹. Within experimental uncertainty, the As($A_{1g}$) mode remains essentially unchanged. Similar trends have been reported in fluorine-doped NdFeAsO system [29]. The hardening of the Fe($B_{1g}$) mode in PrFeAsO$_{1-x}$F$_x$ is attributed to enhanced Fe–Fe interactions resulting from charge transfer the Pr–O/F blocking layers to the Fe–As planes upon F⁻ substitution for O²⁻. This charge transfer increases the electron density in the Fe–As layers, strengthening local Fe–As bonding and favouring superconductivity [30]. In contrast, the observed softening of Pr($A_{1g}$) mode likely reflects lattice relaxation and local strain effects within the Pr–O/F layers caused by the substitution of larger O$^{-2}$ ions with smaller F⁻ ions, introducing local disorder. At higher fluorine concentrations, such disorder may contribute to enhanced scattering and reduced intergranular coupling. Overall, the Raman analysis demonstrate that fluorine doping in PrFeAsO$_{1-x}$F$_x$ primarily modifies the local electronic structure and interlayer charge distribution, while preserving the global crystal symmetry, consistent with the extended stability of the tetragonal superconducting phase.

### c) Transport measurements

The temperature dependence of the electrical resistivity $\rho(T)$ for all prepared Pr1111 samples measured at zero magnetic field is illustrated in Figure 3. The parent compound ($x = 0$) exhibits a pronounced anomaly at 130–140 K, as shown in the inset of Figure 3(a), characteristic of 1111-type iron pnictides. This anomaly arises from a coupled tetragonal-to-orthorhombic structural transition near 135 K, followed by SDW ordering of Fe moments at ~132 K [30]. These features are more clearly resolved in the temperature derivative of resistivity, d$\rho$/d$T$, shown in Supplementary Figure S1 (Section: B1) for $x = 0$ and 0.1. At room temperature, the resistivity of the parent compound is approximately 22 mΩ-cm. Upon cooling, $\rho(T)$ shows a slight increase near the SDW transition, decreases down to ~50 K, and then exhibits an upturn at lower temperatures (inset of Figure 3(b)). A similar resistivity profile is observed for the lightly doped sample $x = 0.1$, with nearly identical anomaly temperature and no evidence of superconductivity (insets of Figures 3(c) and 3(d); Supplementary Figure S1). Both compositions show an additional low-temperature upturn below ~12 K, associated with the antiferromagnetic ordering of Pr$^{3+}$ moments ($T_N^{Pr}$ ≈12 K). This behavior reflects enhanced carrier scattering due to the interplay between Pr magnetic ordering and Fe spin reorientation along the $c$-axis, consistent with earlier reports [13]. These results confirm that the light fluorine substitution ($x \leq 0.1$) is insufficient to induce superconductivity in Pr1111, in agreement with the previous report [23]. For $x = 0.15$, the resistivity displays an approximately linear temperature dependence from room temperature down to ~50 K, followed by a weak upturn and the onset of superconductivity at $T_c^{onset} \approx 20$ K, defined in Figure 4(a). The slight enhancement of $\rho(T)$ just above $T_c$ resembles that observed in the lower-doped samples and likely reflect residual magnetic correlations competing with superconductivity [13]. With further fluorine doping ($x > 0.15$), both the SDW and structural transitions are completely suppressed, coincident with the emergence of bulk superconductivity. Samples with $x = 0.2$ exhibit a linear normal-state resistivity over the entire measured temperature range and a clear superconducting transition. Similar behavior is observed for higher fluorine concentrations ($x = 0.3, 0.35$, and $0.4$), as shown in Figure 3(a). The absence of a systematic trend in the normal-state resistivity magnitude across these doping ranges likely reflects minor variations in the impurity contents, as indicated by XRD (Figure 1). Figure 3(b) summarizes the low-temperature resistivity for $0.15 \leq x \leq 0.4$, while the inset highlights the large low-temperature resistivity of the parent compound. The superconducting onset temperature increases monotonically with fluorine content, reaching ~50 K for $x = 0.4$, demonstrating the progressive stabilization of superconductivity with increasing electron doping in PrFeAsO$_{1-x}$F$_x$.

The resistivity behaviour of highly fluorine-doped PrFeAsO ($x > 0.4$) is shown in Figure 3(c). The samples with $x = 0.5$ and $0.6$ exhibit nearly identical normal-state resistivity and temperature dependence, indicating similar transport characteristics. The sample $x = 0.7$ displays a linear resistivity with a slightly higher magnitude; although a superconducting transition is observed, zero resistivity is not reached, likely due to increased secondary phases identified by XRD. For higher fluorine contents ($x = 0.8$ and $1.0$), the resistivity remains metallic and linear over the entire temperature range, with no evidence of superconductivity down to the lowest measured temperatures, as shown in Figure 3(d). The absence of a systematic trend in the resistivity magnitude and its low-temperature behavior is consistent with increasing structural disorder and impurity contributions. The superconducting onset $T_c$~50.5 K for $x = 0.5$ and increases slightly to 52.3 K for $x = 0.7$, while superconductivity is completely suppressed for $x \geq 0.8$. These observations demonstrate that superconductivity in PrFeAsO$_{1-x}$F$_x$ is stabilized only within the fluorine concentration range $0.15 \leq x \leq 0.7$. Excessive fluorine incorporation leads to phase instability and the dominance of nonsuperconducting impurity phases (e.g., PrAs and PrOF), consistent with the structural analysis. The modest enhancement of $T_c$ for $x = 0.7$ relative to $x = 0.5$ suggests a slight increase in the effective fluorine content within the superconducting lattice, likely below the XRD detection limit, consistent with the nearly unchanged lattice parameters.

The onset ($T_c^{onset}$) and offset ($T_c^{offset}$) of the superconducting transition for the PrFeAsO$_{1-x}$F$_x$ samples are defined in Figure 4(a). The $T_c^{onset}$ corresponds to the temperature at which the resistivity $\rho(T)$ first deviates from the linear normal-state behavior, while $T_c^{offset}$ corresponds to the zero resistivity. The transition width, $\Delta T = T_c^{onset} - T_c^{offset}$, provides a quantitative measure of the superconducting transition broadening across different doping levels in the PrFeAsO$_{1-x}$F$_x$ system. Figures 4(b)–4(d) summarize the evolution of $T_c^{onset}$, $\Delta T$, and the residual resistivity ratio $RRR$ (= $\rho_{300 K}/\rho_{55K}$) as functions of fluorine content ($x$). $T_c^{onset}$ increases rapidly from 20 K for $x$ = 0.15 to 47 K for $x$ = 0.2, reaching 48.3 K for $x$ = 0.3. A slight reduction is observed at $x$ = 0.35 (~ 46 K), followed by a monotonic increase to 50 K at $x$ = 0.4 and a maximum of 52.3 K at $x$ = 0.7. This represents the highest reported $T_c^{onset}$ for the Pr1111 system to date, exceeding previous reports by ~5 K [23]. The observed superconducting transition temperatures of our Pr1111 samples are compared with literature reports [23] in Supplementary Figure S2 (Section: B1), showing good agreement within the previously established fluorine-doping range ($x \leq 0.35$). Based on this evolution, three doping regimes are classified as underdoped (0.15 $\leq x$ <0.3), optimal doped (0.3 $\leq x \leq$ 0.4), and overdoped (0.4 < $x \leq$ 1.0), as presented in Figures 4(b)-(d). The transition width $\Delta T$ (Figure 4(c)) remains nearly constant (~3.5–3.6 K) for $x$ = 0.2–0.35, and then increases almost linearly for higher doping, suggesting reduced grain connectivity at high fluorine contents. The $RRR$ (Figure 4(d)) increases sharply from ~1.5 at $x$ = 0.15 to ~6 for $x$ = 0.2–0.3, reflecting improved crystalline quality, then decreases for $x \geq$ 0.6, consistent the enhanced disorder and impurity scattering. These trends are consistent with the evolution of superconducting behavior and structural characteristics, despite averaging of the intrinsic anisotropy in the polycrystalline nature of the Pr1111 samples [31]. Overall, the systematic evolution of $T_c^{onset}$, $\Delta T$, and $RRR$ with fluorine content confirms that the superconducting behavior in PrFeAsO$_{1-x}$F$_x$ is intrinsic to electronic doping and supports the classification of three doping regimes, consistent with previous studies on *RE*1111 compounds [3]. To analyze the normal-state transport, the zero-field resistivity of PrFeAsO$_{1-x}$F$_x$ samples is fitted in the temperature range 55–300 K using the power-law relation $\rho = \rho_0 + AT^n$, where $\rho_0$ represents the residual resistivity at zero temperature and $A$ is the temperature coefficient. Details of the fitting procedure and associated discussion are provided in the Supplementary Information (Section: B2). Across the entire doping series, the exponent $n$ remains consistently below 2 ($n$ < 2), indicating a robust deviation from conventional Fermi-liquid behavior. Notably, a $T^2$ dependence is absent even near the room temperature, in contrast to several other IBS, such as FeSe$_{1-x}$S$_x$, BaFe$_2$(As$_{1-x}$P$_x$)$_2$, and partially in LaFeAsO$_{1-x}$H$_x$, where Fermi-liquid-like $T^2$ behavior re-emerges at elevated temperatures, as discussed in the Supplementary Information (Section: B2). The persistent non-Fermi-liquid behavior in PrFeAsO$_{1-x}$F$_x$ underscores the presence of unconventional electronic correlations and scattering mechanisms intrinsic to this 1111 system.

### d) Magneto-transport measurements

The temperature dependence of resistivity for the PrFeAsO$_{0.7}$F$_{0.3}$ and PrFeAsO$_{0.65}$F$_{0.35}$ samples under a magnetic field up to 9 T is illustrated in Figures 5(a) and 5(b). The superconducting transition reveals a gradual suppression to lower temperatures and broadens with increasing magnetic field. Following the conventional approach [32], the upper critical field ($H_{c2}$) and irreversibility field ($H_{irr}$) are determined at 90% and 10% of the normal-state resistivity at $T = T_c$, respectively. The resulting $H$-$T$ phase diagrams are shown in the insets of Figures 5(a) and 5(b). The slope of the upper critical field near $T_c$ i.e. $\frac{dH_{c2}}{dT}|_{48\,K}$ is observed to be -6.3 T/K for $x = 0.3$, which is increased to -7.8 T/K ($\frac{dH_{c2}}{dT}|_{46\,K}$ = -7.8 T/K) for the $x = 0.35$, indicating an enhanced field response with increasing fluorine contents. The corresponding slopes of the irreversibility field (d$H_{irr}$/d$T$) are –1.3 T/K and –1.0 T/K for $x = 0.3$ and $x = 0.35$, respectively. Using these parameters, the zero-temperature upper critical fields $H_{c2}(0)$ are estimated according to the Werthamer–Helfand–Hohenberg (WHH) model [33] for single-band and dirty-limit superconductor: $H_{c2}(0) = -0.693 \cdot T_c \cdot \frac{dH_{c2}}{dT}|_{Tc}$; where, $\frac{dH_{c2}}{dT}|_{Tc}$ is the slope of the $H$-$T$ curve at the superconducting $T_c$. The obtained $H_{c2}(0)$ values are ~212 T and 250 T for PrFeAsO$_{0.7}$F$_{0.3}$ and PrFeAsO$_{0.65}$F$_{0.35}$, respectively. These $H_{c2}(0)$ values are comparable to those reported for other 1111-type IBS, such as NdFeAsO$_{0.7}$F$_{0.3}$ ($H_{c2}(0)$~300 T) [34] and SmFeAsO$_{1-x}$F$_x$ ($H_{c2}(0)$~200-400 T) [35], underscoring the robustness of superconductivity in fluorine-doped Pr1111 and highlight its potential for high-field applications. Using the estimated $H_{c2}$ values, we have calculated the coherence length $\xi_{GL}$ using the Ginzburg-Landau (GL) equation near $T_c$ [36],

$$\xi_{GL}(0) = \sqrt{\frac{\Phi_0}{2\pi\mu_o H_{c2}(0)}}$$

where $\Phi_0$ is the magnetic flux quantum h/2e, $\mu_o$ is the vacuum permeability $4\pi \cdot 10^{-7}$ H/m. This yields $\xi_{1GL}(0) = 1.23$ nm and $\xi_{2GL}(0) = 1.15$ nm for the PrFeAsO$_{0.7}$F$_{0.3}$ and PrFeAsO$_{0.65}$F$_{0.35}$ samples, respectively. Such short coherence lengths, together with the large upper critical fields, place Pr1111 in the dirty-limit regime of superconductivity. Furthermore, by numerically solving the equation $g(\lambda_{G_L}) = 0$, expressed as $g(\lambda_{G_L}) = H_{c1} - \frac{\Phi_0}{4\pi\lambda_{GL}^2}[ln\,\kappa + 0.5] = 0$ [36] where $\kappa = \frac{\lambda_{G_L}}{\xi_{GL}}$, we have calculated the penetration depth $\lambda_{1GL}(0) = 142$ nm and $\lambda_{2GL}(0) = 165$ nm for the sample $x = 0.3$ and $0.35$, respectively by assuming weak-pinning regime and $H_{irr}(0) \sim H_{c1}(0)$ [37]. The resulting Ginzburg-Landau parameter $\kappa = \lambda/\xi$ is also calculated, yielding $\kappa_1 \sim 115$ and $\kappa_2 \sim 143$ for $x = 0.3$ and $0.35$, respectively. These values ($\kappa \gg 1/\sqrt{2}$) confirm the type-II nature of superconductivity in F-doped PrFeAsO and are consistent with earlier reports on *RE*1111 systems [38], [35]. The short coherence length ($\xi$) can be attributed to the large superconducting gap $\Delta$, as described by a relation [36] $\xi \sim \frac{\hbar v_F}{\Delta}$ where $v_F$ is the Fermi velocity, a hallmark of high-$T_c$ superconductors [39].

In high $T_c$ superconductors, the superconductivity can be suppressed by both orbital and spin-paramagnetic pair-breaking mechanisms. While the WHH model accounts primarily for orbital effects, spin-paramagnetic contributions are known to be significant in IBS [32]. The Pauli-limiting field, $H_p = 1.84 \cdot T_c$, is estimated to be 88.3 T for $x = 0.3$ and 84.6 T for $x = 0.35$, yielding Maki parameters $\alpha\,(=\sqrt{2}\,\frac{H_c^{orb}(0)}{H_P}) = 3.3$ and $2.9$, respectively. These large Maki parameter values ($\alpha > 1.8$) suggest a dominant role of Pauli-limiting effects and suggest the possible involvement of unconventional superconducting states arising from the competition between orbital and paramagnetic pair breaking, such as the Fulde–Ferrell–Larkin–Ovchinnikov (FFLO)-like or spin-triplet components [40]. The prominence of spin-paramagnetic effects, together with the observed non-Fermi-liquid normal-state transport and short coherence lengths, supports a multiband superconducting scenario with unconventional pairing in PrFeAsO$_{1-x}$F$_x$. This interpretation is consistent with recent spectroscopic and NMR studies reporting multigap superconductivity and unconventional pairing symmetry in this system [41].

A key parameter governing the performance of type-II superconductors is the critical current density $J_c$, which is controlled by the competition between vortex pinning forces and the Lorentz force acting on vortices. When pinning dominates, vortices undergo thermally activated flux creep, leading to an Arrhenius-type resistivity behavior, whereas stronger Lorentz forces drive flux flow and dissipation [37]. Figures 5(c) and 5(d) display the Arrhenius plots of the resistivity as a function of inverse temperature (1/$T$) for PrFeAsO$_{0.7}$F$_{0.3}$ and PrFeAsO$_{0.65}$F$_{0.35}$, respectively. The low-temperature linear regions are fitted by using $\rho =$ $\rho_0 \, exp\left(-\frac{U_0(T,B)}{k_B T}\right)$, where the normal-state resistivity at 55 K is used as $\rho_0$. The extracted activation energy $U_0$ decreases systematically with increasing magnetic field $H$, as shown in the insets of Figures 5(c) and 5(d), consistent with thermally activated vortex motion. The magnitude of $U_0$ is comparable to that reported for PrFeAsO$_{0.6}$F$_{0.12}$ [17], suggesting similar pinning characteristics across the Pr1111 system. The field dependence of the $U_0$ typically follows a power-law behaviour, $U_0 \sim H^{-\eta}$, with two distinct regimes, as shown in the insets of Figure 5(c) and 5(d). For $H < 4$ T, the exponents are $\eta = 0.30$ and $0.41$ for $x = 0.3$ and $0.35$, respectively, while at higher fields, $\eta$ increases to 0.69 and 0.55. This crossover signifies a transition from single-vortex pinning at low fields to collective pinning at higher fields, in agreement with observations in other 1111-type superconductors such as NdFeAsO$_{0.7}$F$_{0.3}$ [34] and SmFeAsO$_{0.9}$F$_{0.1}$ [42]. The relatively low activation energies ($U_0 \sim 500$ K), compared with MgB$_2$ ($U_0 \sim 10^4$ K) [43], reflect intrinsically weaker vortex pinning in IBS. Consequently, the critical current density in PrFeAs(O,F) is likely governed by weak collective pinning arising from atomic-scale disorder or point defects [44], as commonly observed in the 1111 family.

### (e) Magnetization measurements

To further confirm bulk superconductivity and the Meissner effect, magnetization measurements were performed on selected PrFeAsO$_{1-x}$F$_x$ samples in the temperature range 2–60 K under an applied magnetic field of 20 Oe. Figure 6 presents the temperature dependence of the normalized magnetic moment ($M/M_{5K}$) measured under both zero-field-cooled (ZFC) and field-cooled (FC) conditions for the samples with $x = 0.25, 0.3, 0.35, 0.4,$ and $0.5$. Additionally, the corresponding data for the sample $x = 0.2$ are presented in the supplementary figure S7(a) (Section: B3), confirming a superconducting transition at ~46.5 K with an almost single-step feature. The $x = 0.25$ sample exhibits a characteristic double-step transition with an onset superconducting transition temperature ($T_c$) of 39.4 K, consistent with the transport measurements. Such behaviour is commonly attributed to weak intergranular coupling and the coexistence of inter- and intragrain superconducting responses [45] in polycrystalline 1111 compounds. With increasing fluorine content, the onset $T_c$ increases to 46.1 K for $x = 0.3$ and remains nearly unchanged for $x = 0.35$ and 0.4, before reaching ~49 K for $x = 0.5$, in agreement with the resistivity data. Notably, the $x = 0.35$ sample exhibits a sharp single-step transition, indicating enhanced grain connectivity near optimal doping, whereas other compositions retain the double-step feature typical of 1111-type superconductors [28]. The magnetically determined the onset $T_c$ values are systematically ~1 K lower than those obtained from the resistivity measurements, consistent with previous reports on IBS [10]. Magnetic hysteresis loops measured at 5 K for $x$ = 0.2, 0.25, 0.3, 0.35, 0.4 and 0.5 are used to estimate the $J_c$. The detailed analysis and discussion are provided in the Supplementary Information (Section: B3). The extracted $J_c$ values for these polycrystalline PrFeAsO$_{1-x}$F$_x$ samples are on the order of $10^3$ A/cm$^2$, which is approximately two to three orders of magnitude lower than those reported for single-crystal PrFeAsO$_{0.60}$F$_{0.35}$ (~$10^5$ A/cm$^2$) [38]. This discrepancy highlights the dominant role of microstructural inhomogeneity and weak intergranular coupling in limiting current transport in bulk polycrystalline Pr1111 samples, emphasizing the need for improved synthesis and densification strategies to enhance grain connectivity and vortex pinning. The corresponding pinning force ($F_p$) for these samples are also discussed in the Supplementary Information (Section: B4).

e) **Specific heat measurements**

The superconducting nature of the F-doped PrFeAsO system is further confirmed by the specific heat measurements performed on bulk PrFeAsO$_{0.65}$F$_{0.35}$. Figure 7(a) illustrates the temperature dependence of specific heat plotted as *C/T* over the range 2-300 K under the applied magnetic fields of 0 T and 1 T, and 2-100 K under the fields up to 9 T. The normal-state behaviour of PrFeAsO$_{0.65}$F$_{0.35}$ is consistent with that of other iron-pnictide superconductors. A pronounced low-temperature anomaly near 10 K is observed and attributed to the magnetic ordering of Pr$^{3+}$ ions, as reported previously for 1111 family [46], [47]. As shown in Figure 7(b), this anomaly shifts to higher temperatures with increasing magnetic field, indicating a crystalline electric field (CEF)–induced Schottky contribution arising from the splitting of the Pr$^{3+}$ ground-state multiplet [48]. The specific heat data in the temperature range 7-35 K are analyzed using a slightly modified combined electronic, Debye, Schottky model, following a model previously proposed for PrFePO [49] and given below. This model, as shown in Figure 7(c), effectively captures both the Pr$^{3+}$ magnetic peak and the underlying electronic and phononic background, confirming the coexistence of superconductivity and localized 4f magnetism in PrFeAsO$_{0.65}$F$_{0.35}$, a hallmark of rare-earth 1111 iron pnictides.

$$C = \gamma T + C_D(T, \theta_D) + C_{sch}(T)$$

where, $C_D(T, \theta_D) = 9R \left(\frac{T}{\theta_D}\right)^3 \int_0^{\frac{\theta_D}{T}} \frac{x^4 \, dx}{(e^x - 1)^2}$, where $x = \frac{\hbar \omega}{k_B T}$ employed by [50] and, $C_{sch}(T) = R \Delta^2 \frac{g_0}{g_1} \frac{\exp \Delta}{[1 + (g_0/g_1) \exp \Delta]^2}$ where, $\Delta = \frac{\delta}{T}$, employed by [49]

Where $R$ is the molar gas constant, $C_D(T,\theta_D)$ [50] represents the Debye's function, $\gamma$ is the electronic specific heat coefficient. From the low-temperature fit, the $\gamma$ is determined to be 33 mJ mol$^{-1}$ K$^{-2}$, in good agreement with previous reports [13], [21]. The extracted Debye temperature $\theta_D$ is 294 K (for low-temperature region), consistent with values typical for iron pnictides [51]. The relatively large $\gamma$ compared to La-1111 ($\gamma = 3.7$ mJ mol$^{-1}$ K$^{-2}$) [52] reflects the additional low-temperature contribution from the Pr-related Schottky anomaly, as evident from the peak feature in figure 7(c). The Schottky contribution $C_{sch}$ describes the specific heat anomaly of a two-level system, characterized by an energy separation $\delta$ (in K) and associated degeneracies $g_0$ (ground state) and $g_1$ (first excited state). Following the model proposed for PrFePO [49], the Schottky analysis yields an energy gap $\Delta \sim 26$ K (2.24 meV), consistent with a low-lying CEF excitation, possibly corresponding to a forbidden transition between singlet states in the F-doped system [21]. The amplitude of the CEF-related contribution, $g$, is calculated to be 0.39, implying that only about 39% of the expected entropy (Rln2 ~5.76 J mol$^{-1}$ K$^{-1}$) is released from the ground-state doublet, resulting in a recovered entropy of ~2.25 J mol$^{-1}$ K$^{-1}$. Such a reduced value of entropy implies a notable hybridization between the itinerant Fe 3d electrons and the localized Pr 4f electrons, signifying a subtle interplay between magnetic and electronic degrees of freedom in the PrFeAsO$_{1-x}$F$_x$ system [13].

The high-temperature specific heat $C$ data (50–200 K) are analyzed using a simplified Debye model [50]:

$$C = \gamma T + A_D C_D(T,\theta_D)$$

The fitted parameters are presented in Figure 7(d), yielding an electronic specific heat coefficient $\gamma = 169$ mJ mol$^{-1}$ K$^{-2}$, and a Debye temperature $\theta_D = 285$ K. The relatively high $\gamma$ value is likely overestimated due to the neglect of magnetic contributions from the Pr$^{3+}$ ions. Such an overestimation in $\gamma$ has also been reported in previous studies, where variations in $\gamma$ were attributed to the complex interplay between localized 4f moments and conduction electrons [51], [48]. Therefore, the $\gamma$ value derived from the low-temperature fit ($\gamma = 33$ mJ mol$^{-1}$ K$^{-2}$, obtained in the 7-35 K range) is adopted as the intrinsic Sommerfeld coefficient for PrFeAsO$_{0.65}$F$_{0.35}$.

To examine the superconducting contribution, the specific heat measured at 0 T, ($C(0\ T)$), is subtracted from that data collected at 9 T, ($C(9\ T)$), and the resulting [$C(0\ T) - C(9\ T)$]/$T$ with temperature dependence is presented in Figure 8(a). A clear anomaly corresponding to the superconducting transition is observed at ~44.5 K, approximately 1 K lower than the transition temperature obtained from the resistivity measurements for the $x = 0.35$ sample. This small $T_c$ deviation may arise from the filamentary superconductivity or weak intergranular coupling effects. The obtained specific heat jump, $\Delta C/\gamma T_c$ is found to be 0.14, which is well below the weak-coupling BCS limit value of 1.43, consistent with other members of the 1111 family [53]. The absolute magnitude of the jump $\Delta C/T$, $[C(0\,\text{T}) - C(9\,\text{T})]/T$, is 4.66 mJ mol$^{-1}$ K$^{-2}$, is relatively small across the $T_c$, implying a large Ginzburg number $G_i$ ($\sim \left(\frac{k_B T_C}{\Delta C \xi^3}\right)^2$) and hence strong thermal fluctuations near $T_c$. The electronic density of states at the Fermi level, $N(E_F) = \frac{3\gamma}{\pi^2 k_B^2 (1+\lambda_{e-ph})}$, is estimated to be approximately 6.95 eV$^{-1}$ f.u.$^{-1}$, comparable to values reported for other 1111-type superconductors [51]. The electron-phonon coupling strength, $\lambda_{e-ph}$, as determined by the McMillan formula [54], is reported to be $\sim 1.03$, suggesting moderately strong coupling. Similarly, we have subtracted the specific heat at 0 T ($C(0T)$) from that measured at various applied magnetic fields ($\mu_0 H$ = 1 to 9 T), and plotted the resulting specific heat jump $\Delta C/T$ curves in Figure 8(b). In contrast to most 1111 compounds, where the specific heat jump decreases with increasing magnetic field [51], the present samples exhibit an opposite trend. Although the origin of this behaviour remains unclear, it may be influenced by nonsystematic impurity phases (Figure 1), since both magnetic and nonmagnetic impurity scattering can strongly modify thermodynamic responses in multiband superconductors [55]. From the field dependence of the transition midpoint, we have estimated the slope d$H_{c2}$/dT at $T = T_c$, yielding a value of approximately -3.37 T/K and the obtained $H_{c2}(0) \sim 104$ T according to the WHH model [33]. This value is lower than that derived from magnetotransport measurements, reflecting granularity effects and highlighting the need for improved sample homogeneity and intergrain connectivity in bulk F-doped PrFeAsO.

## IV. DISCUSSION

Previous studies on electron-doped 1111-type iron pnictides have largely been restricted to limited substitution ranges, including LaFeAsO$_{1-x}$F$_x$ ($0 < x \leq 0.2$), LaFeAsO$_{1-x}$H$_x$ ($0 < x \leq 0.5$), SmFeAsO$_{1-x}$F$_x$ ($0 < x \leq 0.2$), PrFeAsO$_{1-x}$F$_x$ ($0 < x \leq 0.35$) or CeFeAsO$_{1-x}$F$_x$ ($0 < x \leq 0.2$). These systems typically exhibit one or two superconducting domes as a function of dopant concentration [56], [30], [11], [23]. While fluorine substitution is generally limited to ~20–25%, hydrogen doping—achieved under high-pressure synthesis—can extend to ~50–60%; however, the interpretation of superconductivity in H-doped 1111 compounds remains controversial due to difficulties in quantifying hydrogen content and the possible role of oxygen deficiency. By combining systematic transport, magnetization, and specific heat measurements, we construct a comprehensive electronic phase diagram for PrFeAs(O,F), shown in Figure 9, covering the full nominal fluorine range $0 \leq x \leq 1$. The superconducting $T_c^{onset}$ are extracted from transport, magnetization, and specific heat measurements, while the spin-density-wave transition, represented by the Fe magnetic ordering temperature ($T_N^{Fe}$), and the Pr magnetic ordering temperature ($T_N^{Pr}$) are derived from resistivity anomalies. The parent ($x = 0$) and the lightly doped ($x = 0.1$) samples exhibit SDW features near 130-140 K, followed by Pr magnetic ordering at $T_N^{Pr} \approx 12$ K. Bulk superconductivity first emerges at $x = 0.15$, where weak residual SDW signature coexists with superconductivity, indicating a narrow region of competing electronic orders. With increasing fluorine content, the SDW order is fully suppressed, and a robust superconducting phase develops, forming a broad dome extending from the underdoped regime ($x = 0.15$, $T_c = 20$ K) to the overdoped limit ($x = 0.7$), where a maximum $T_c$ of 52.3 K is achieved. Beyond this range ($x = 0.8$ and $x = 1$), superconductivity is absent and the system stabilizes in a nonsuperconducting cubic phase, as shown in Figure 9. The superconducting $T_c$ values obtained from magnetization for $0.20 \leq x \leq 0.50$ are systematically 1 K lower than those derived from transport measurements, while the specific heat transition for $x = 0.35$ occurs at $T_c = 44.5$ K, slightly below the resistive-derived $T_c = 46$ K. A comparison with the phase diagram reported by Rotundu *et al.* [23] for F-doped PrFeAsO, which showed a maximum $T_c$ of 47 K at $x = 0.25$ (Supplementary Figure S2), reveals that our results significantly extend both the accessible fluorine-doping range and the superconducting dome. To our knowledge, this work provides the first continuous electronic phase diagram spanning the full nominal fluorine composition in a 1111-type iron pnictide. The emergence of a broad superconducting dome with a higher optimal $T_c$ underscores the critical role of systematic doping control in elucidating the interplay between magnetism and superconductivity and in constraining pairing mechanisms in IBS.

## V. CONCLUSIONS

A complete series of PrFeAsO$_{1-x}$F$_x$ ($0 \leq x \leq 1$) compounds was successfully synthesized, enabling construction of the first comprehensive electronic phase diagram spanning the full fluorine-doping range in the *RE*1111 family. A systematic contraction of the lattice parameters and unit-cell volume with increasing fluorine content ($x$) confirms effective substitution at the oxygen sites. This result is further supported by Raman spectroscopy, which reveals systematic phonon renormalization consistent with charge transfer between the PrO and FeAs layers. Transport and magnetic measurements establish bulk superconductivity over a wide doping range $0.15 \leq x \leq 0.7$, with a maximum $T_c \approx 52.3$ K —the highest reported to date for the Pr1111 system. At full substitution ($x = 1$), the tetragonal superconducting phase is replaced by a cubic, nonsuperconducting phase, indicating a fundamental structural and electronic instability at extreme fluorine concentrations. The normal-state resistivity of the superconducting compositions exhibits pronounced non–Fermi-liquid behavior, pointing to enhanced spin and/or orbital fluctuations. The large upper critical fields and Maki parameters exceeding unity suggest a dominant role of spin-paramagnetic pair-breaking effects. The analysis of thermally activated flux flow reveals a crossover from single-vortex pinning at low fields to collective pinning at higher magnetic fields, while the reduced specific-heat jump near $T_c$ reflects strong superconducting fluctuations. These results support the multiband nature of the superconductivity in Pr1111 system, consistent with the broader class of IBS. The resulting phase diagram reveals a well-defined superconducting dome comprising the underdoped ($0.15 \leq x < 0.3$), optimal doped ($0.3 \leq x \leq 0.4$), and overdoped ($0.4 < x \leq 0.7$) regimes. Collectively, this work resolves a long-standing gap in the electronic phase diagram of the 1111 family and demonstrates that optimized synthesis conditions can substantially extend the fluorine solubility limit in $RE$1111 compounds. Our findings provide important insights into the interplay between electronic correlations, magnetism, and superconductivity in IBS.

## Acknowledgments

The work was funded by SONATA-BIS 11 project (Registration number: 2021/42/E/ST5/00262) sponsored by National Science Centre (NCN), Poland. SJS acknowledges financial support from National Science Centre (NCN), Poland through research Project number: 2021/42/E/ST5/00262.

## CRediT authorship contribution statement

**Priya Singh:** Writing – review & editing, Writing – original draft, Investigation, Formal analysis, Data curation. **Konrad Kwatek:** Formal analysis, Investigation, Resources, Data curation, Writing – review & editing. **Tatiana Zajarniuk:** Data curation, Investigation,

Resources, Writing – review & editing. **Taras Palasyuk:** Writing – review & editing, Formal analysis, Investigation, Data curation. **Cezariusz Jastrzębski:** Writing – review & editing, Resources, Data curation. **Andrzej Szewczyk:** Writing–review & editing, Resources, Data curation. **Shiv J. Singh:** Writing–review & editing, Writing – original draft, Visualization, Validation, Supervision, Software, Resources, Methodology, Investigation, Funding acquisition, Formal analysis, Conceptualization.

**Figure 1:** Powder X-ray diffraction (XRD) patterns of PrFeAsO$_{1-x}$F$_x$ samples are presented for (**a**) $x =$ 0, 0.1, 0.15, 0.2, 0.25, 0.3, 0.35, 0.4 and (**b**) $x =$ 0.5, 0.6, 0.7, 0.8, 1.0. (**c**) Rietveld refinement of the room temperature XRD pattern for the sample $x = 0.3$, showing the experimental data, calculated profile, and the corresponding PrFeAsO (Pr1111) phase pattern, and the difference curve. The variation of (**d**) lattice parameter '*a*', (**e**) lattice parameter '*c*', and (**f**) lattice volume '*V*' as a function of the nominal fluorine content (*x*). The possible error bars in lattice parameters and volume are included in panels (d)–(f). For comparison, the lattice parameters and unit cell volumes from Kodama et al. [18] and Meena et al. [26] are also included in panels (d)–(f).

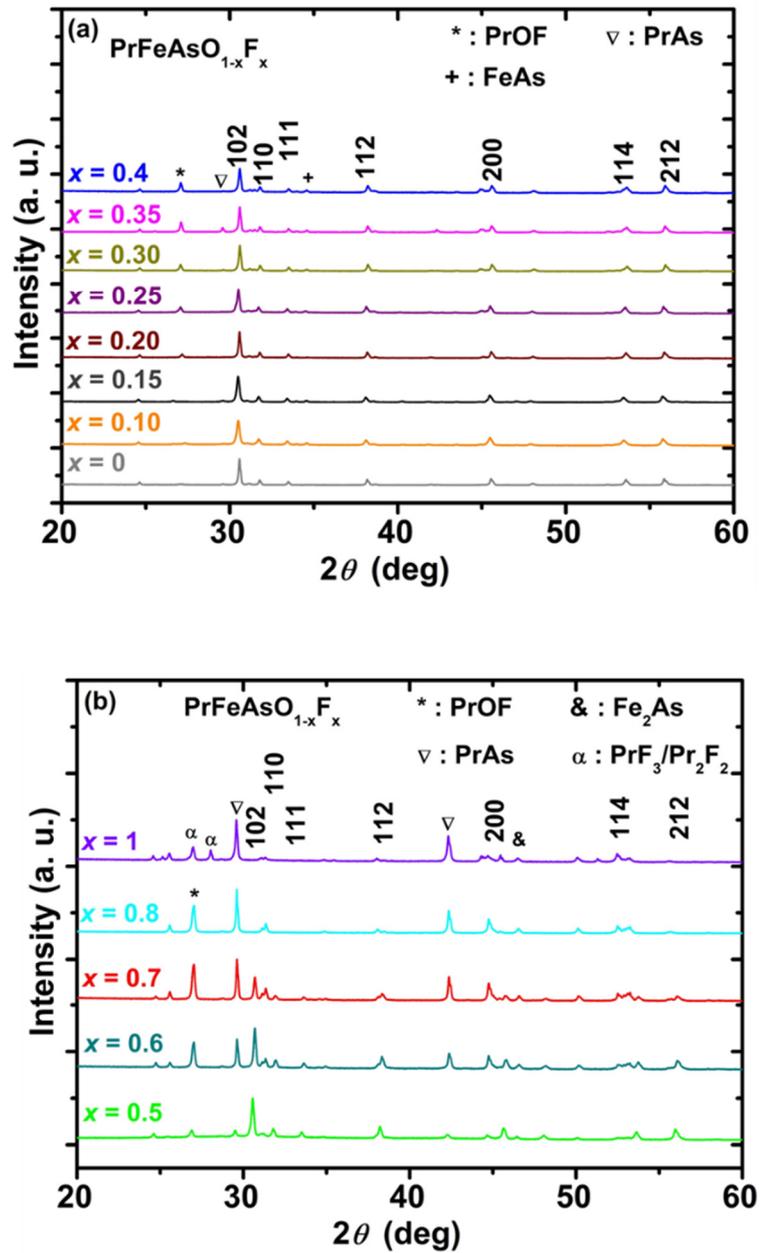

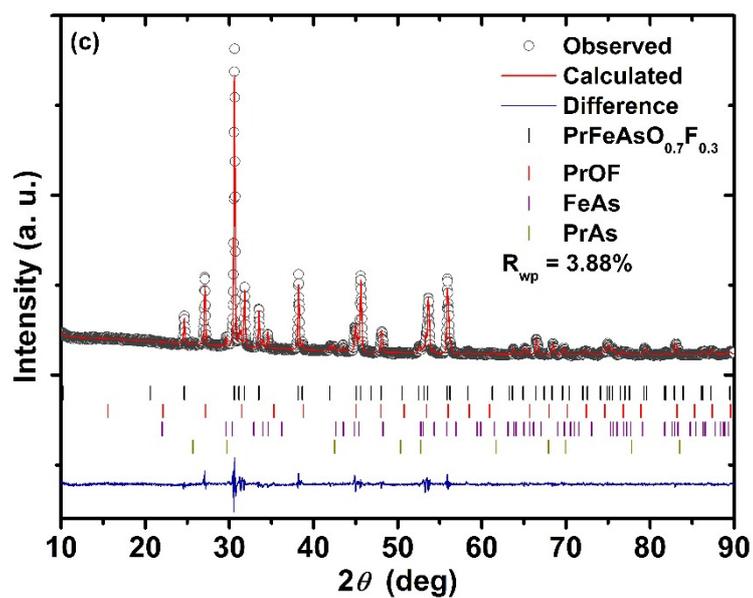

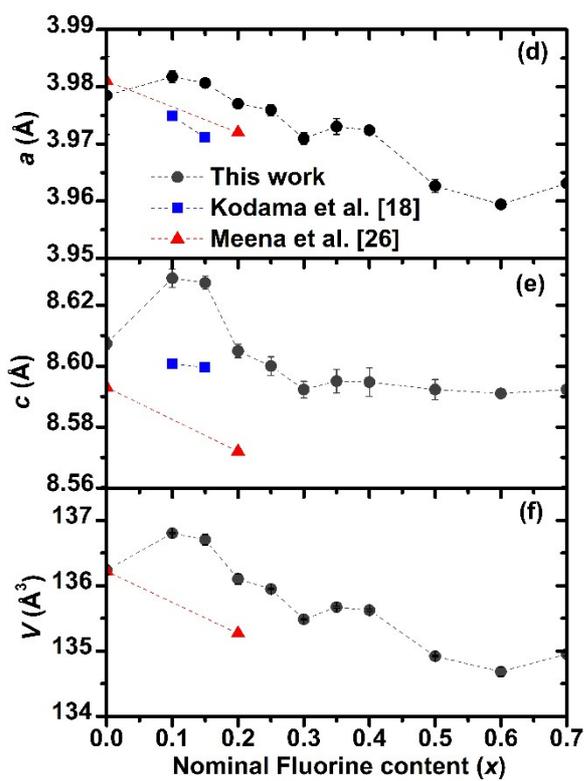

**Figure 2:** Raman scattering spectra of fluorine-doped PrFeAsO$_{1-x}$F$_x$ with varying fluorine contents $X_F$ are presented. **(a)** Representative Raman spectra of the pristine compound (0%; $x = 0$) are compared to those of the doped samples with fluorine contents of 30% ($x = 0.3$) and 50% ($x = 0.5$). The spectra are offset vertically for clarity. The assignment of detected signals related to lattice vibrations is shown for the spectrum of pristine material. Deconvolution of the experimental spectra with the Lorentz function is illustrated with the green lines. Total fits of Lorentz model to experimental data are shown by red lines. Vertical black lines are provided as a guide to the eye. **(b)** The evolution of Raman peak positions as a function of fluorine content is depicted. Experimental data points are shown as circles with corresponding error bars, while dashed lines serve as visual guides to highlight the systematic peak shifts induced by fluorine substitution.

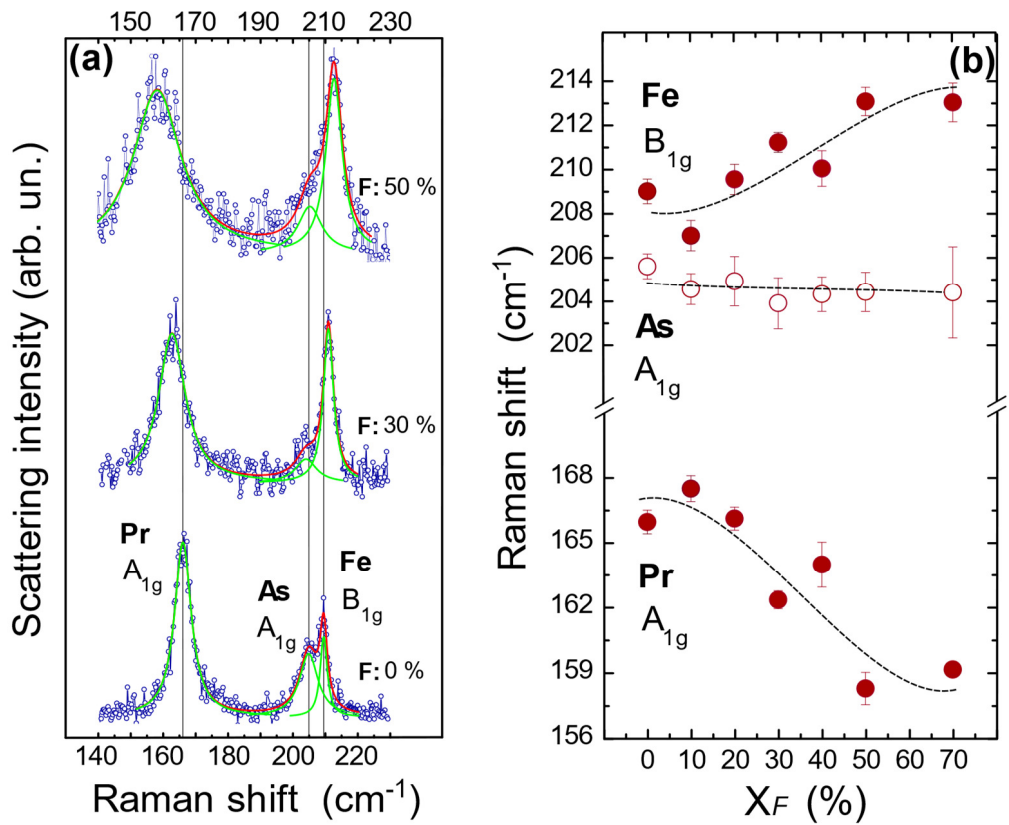

**Figure 3:** The temperature dependence of the resistivity $\rho$ for PrFeAsO$_{1-x}$F$_x$ samples with $x$ = 0.15 to 0.4 is shown (**a**) up to the room temperature, and (**b**) in the low-temperature region (below 60 K). The inset of figures (a) and (b) depict the temperature dependence of the resistivity of parent sample $x = 0$ over the full and low temperature regions, respectively. The temperature dependence of the resistivity $\rho$ of PrFeAsO$_{1-x}$F$_x$ with $x$ = 0.5 to 1.0 is presented (**c**) across the entire measured temperature range (7-300 K) and (**d**) in the low-temperature region (7-60 K). The temperature dependency of resistivity for the sample $x = 0.1$ is depicted in the inset of figures (c) and (d) over the full and low-temperature regions, respectively. The Log scale of Figure 3(c) and 3(d) are depicted in the Supplementary Figure S3(a) and S3(b) (Section: B1).

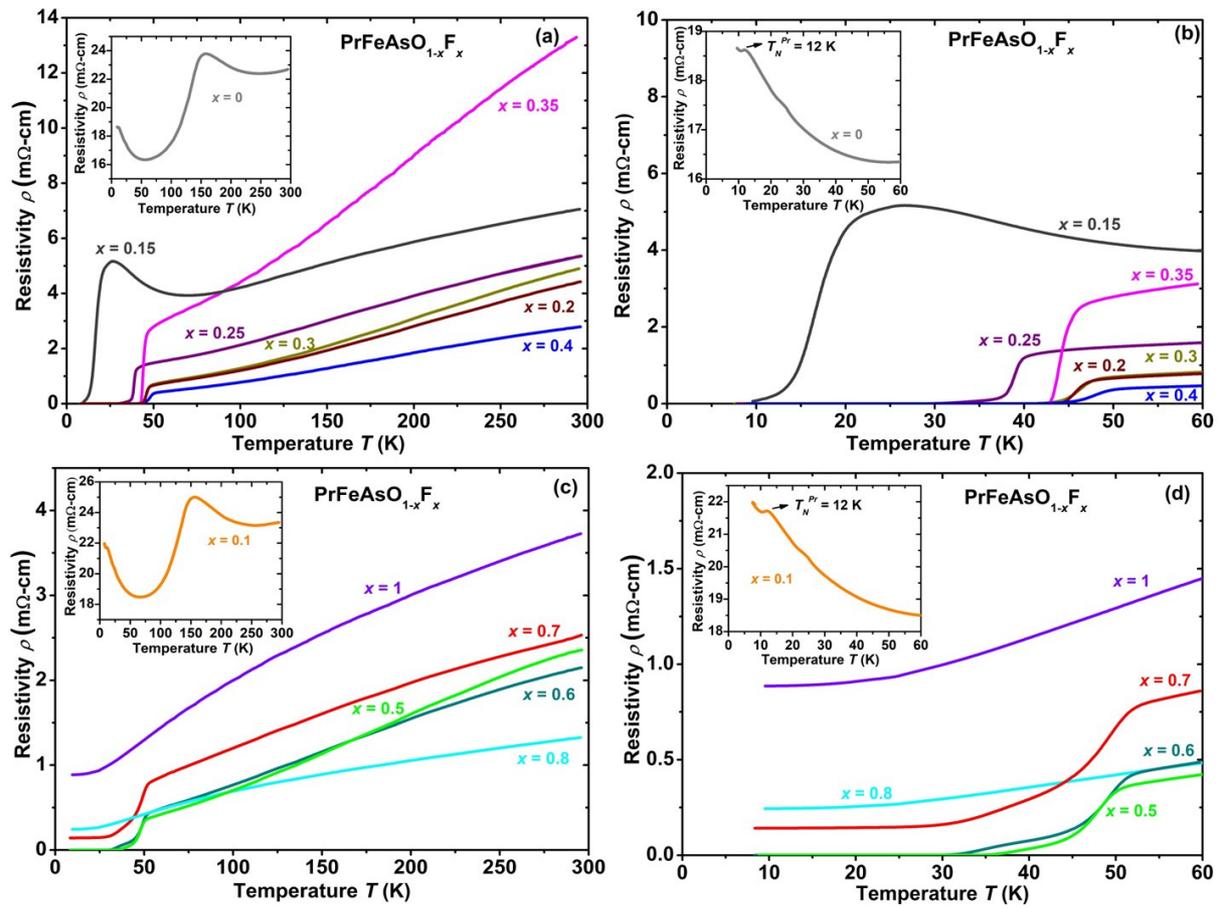

**Figure 4: (a)** Low temperature dependence of resistivity for the PrFeAsO$_{0.7}$F$_{0.3}$ sample, illustrating the method used to determine the onset ($T_c^{onset}$) and offset ($T_c^{offset}$) of the superconducting transition. The variation of **(b)** the onset critical transition temperature ($T_c^{onset}$), **(c)** transition width ($\Delta T = T_c^{onset} - T_c^{offset}$), and **(d)** residual resistivity ratio $RRR$ (= $\rho_{300K}/\rho_{55K}$) as a function of nominal fluorine content ($x$) for PrFeAsO$_{1-x}$F$_x$.

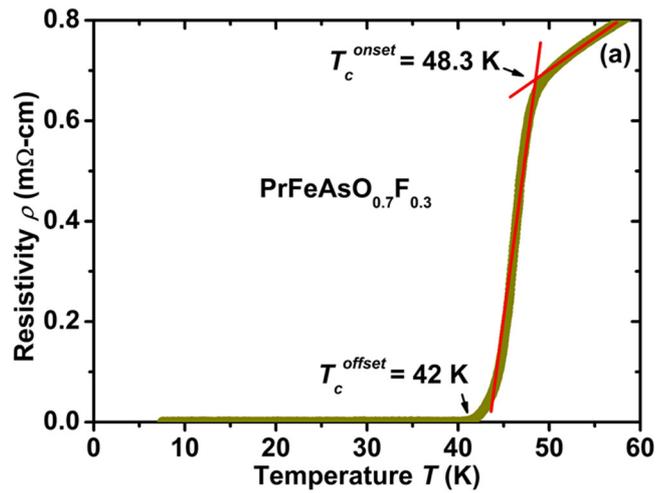

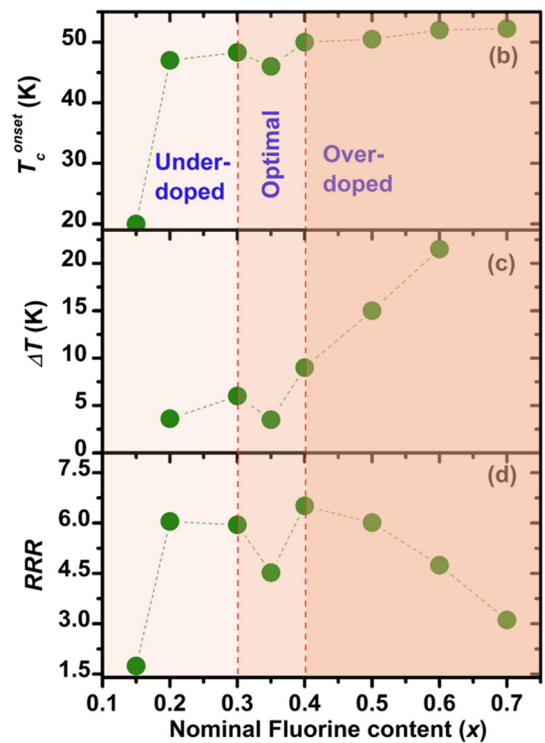

**Figure 5:** The variation of resistivity with temperature up to 55 K under the applied magnetic fields ranging from 0 T to 9 T is presented for (**a**) PrFeAsO$_{0.7}$F$_{0.3}$ and (**b**) PrFeAsO$_{0.65}$F$_{0.35}$ respectively. The inset figure of (a) and (b) illustrate the corresponding *H-T* phase diagrams, constructed using 90% and 10% of normal-state resistivity just above the superconducting transition temperature to estimate the slopes d$H_{c2}$/d$T$ and d$H_{irr}$/d$T$. The logarithmic variation of normalized resistivity $\rho/\rho_{55K}$ as a function of the inverse of temperature $1/T$ under the applied magnetic fields of 1-9 T for (**c**) PrFeAsO$_{0.7}$F$_{0.3}$ and (**d**) PrFeAsO$_{0.65}$F$_{0.35}$, respectively. The inset figure of (c) and (d) display the plot of calculated thermally activated flux flow activation energy $U_0$ as function of the applied magnetic field for $x$ = 0.3 and 0.35, respectively.

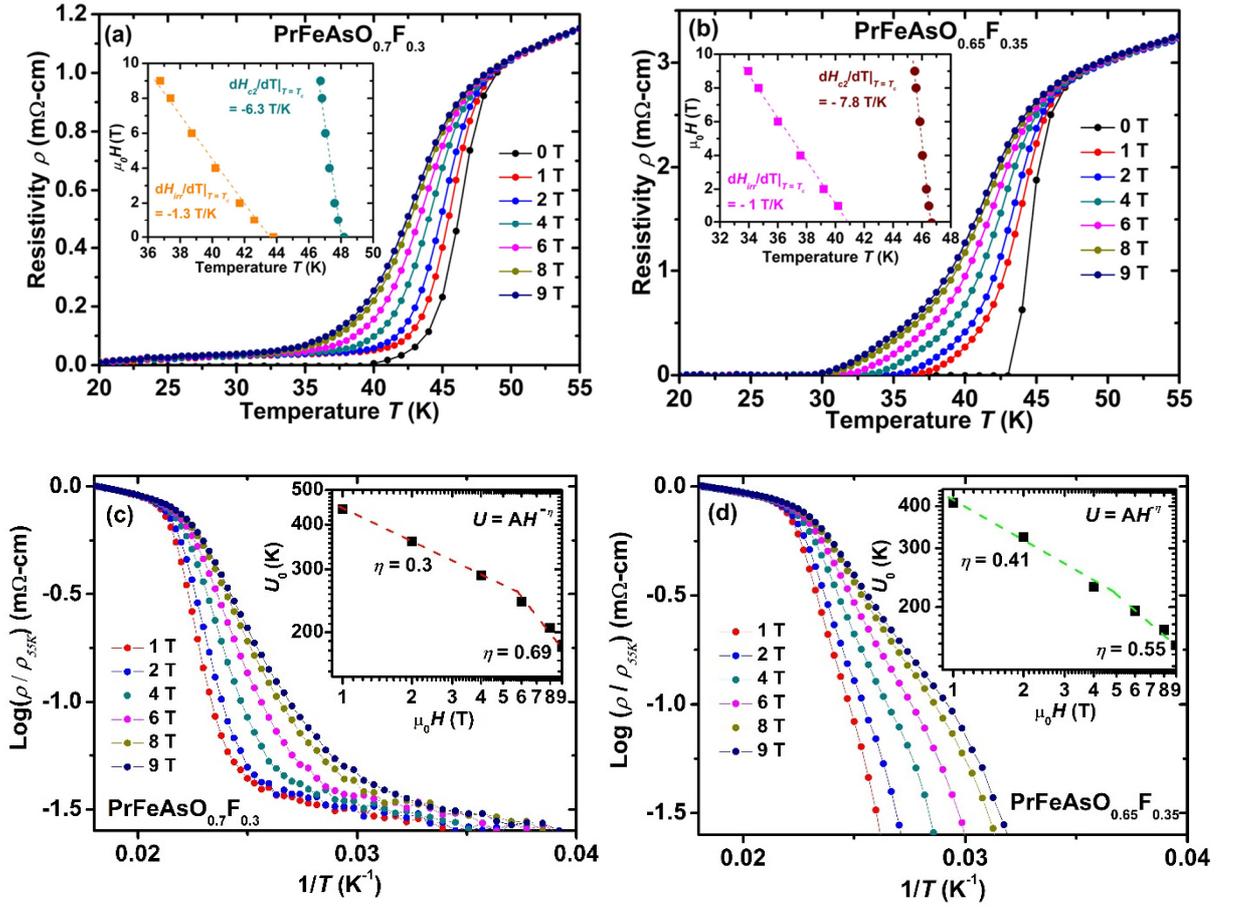

**Figure 6:** The temperature dependence of the normalized magnetic moment $M/M_{5K}$ for PrFeAsO$_{1-x}$F$_x$ samples with $x$ = 0.25, 0.3, 0.35, 0.4 and 0.5, measured under an applied magnetic field of 20 Oe in both zero-field cooling (ZFC) and field cooling (FC) modes.

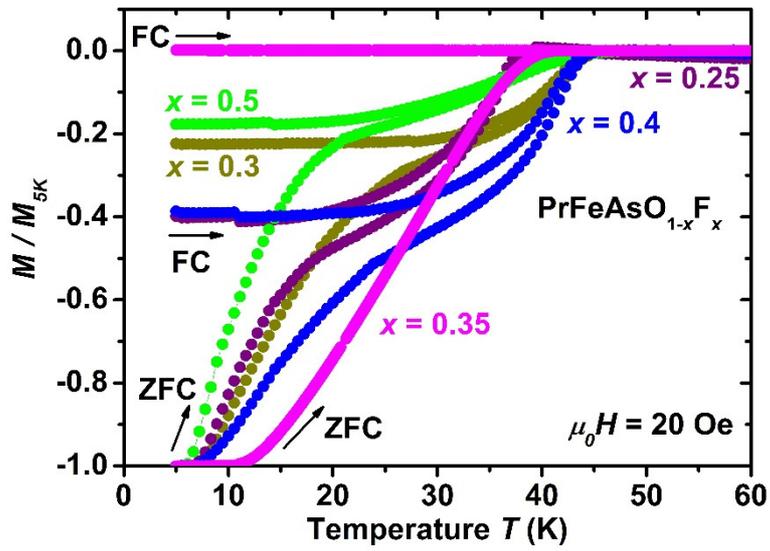

**Figure 7:** (a) The temperature variation of the specific heat $C/T$ over the full measured temperature range under different applied magnetic fields $\mu_0 H = 0$ T to 9 T for PrFeAsO$_{0.65}$F$_{0.35}$. (b) The plot of the specific heat $C/T$ as a function of temperature $T$ in the low-temperature region (2-20 K) under the applied magnetic field $\mu_0 H = 0$ T to 9 T, indicating the suppression of the Pr-related peak with the increasing magnetic field. (c) The variation of the specific heat $C/T$ with temperature $T$ in the low temperature range below 50 K at 0 T, showing the fitting performed in the range 7-35 K (green line) by considering the Schottky anomaly, electronic and lattice contributions to the specific heat. (d) The variation of the specific heat $C$ with temperature $T$ is depicted from 2 K to 200 K under zero magnetic field, with the high-temperature (50–200 K) region fitted (green line) using the combined electronic and Debye lattice contributions.

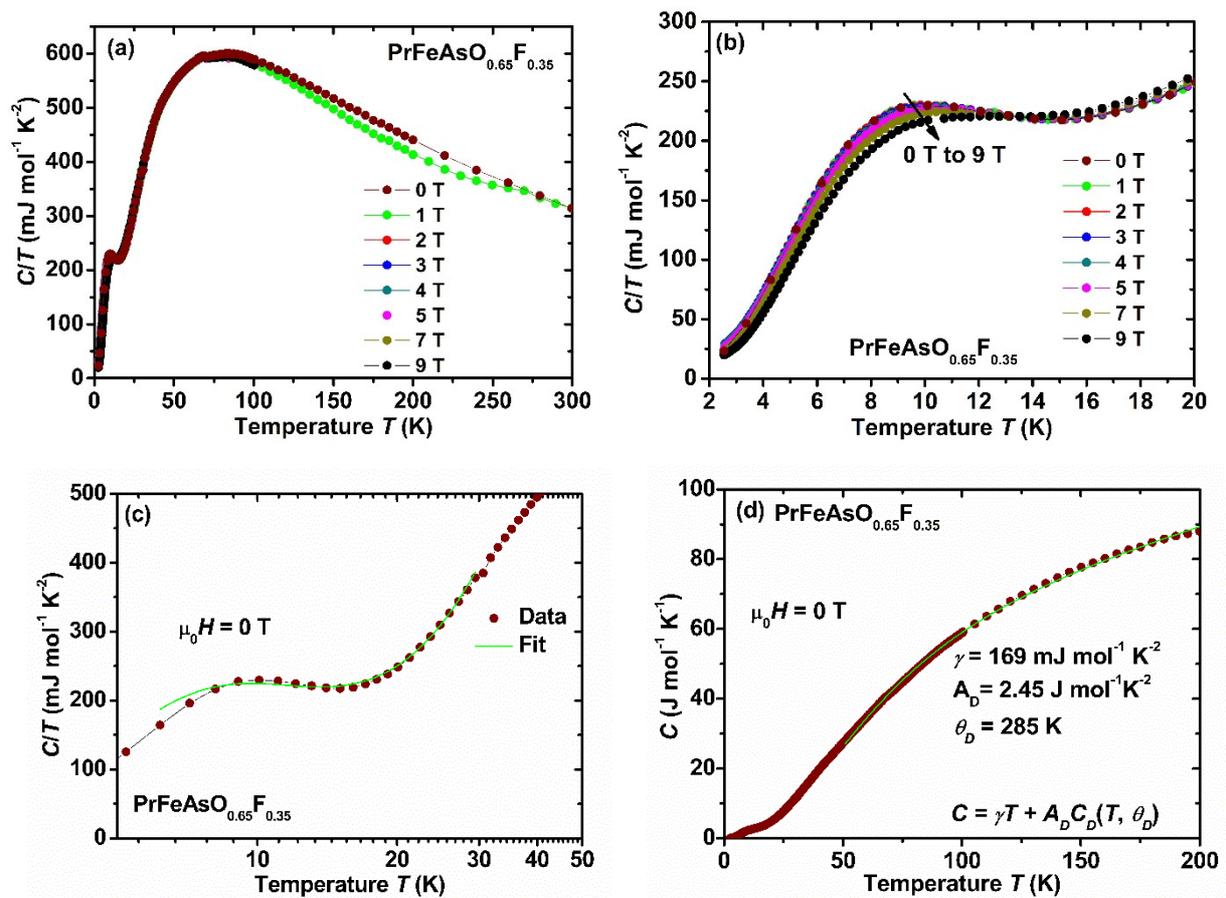

**Figure 8:** **(a)** The temperature dependence of the difference in the specific heat data, $[C(0\,T)-C(9\,T)]/T$, obtained by subtracting the data measured at 9 T ($C(9\,T)$) from that at zero field ($C(0\,T)$), illustrating the superconducting transition $T_c$ for $PrFeAsO_{0.65}F_{0.35}$. **(b)** The temperature variation of the specific heat jump, $\Delta C/T$, obtained by subtracting the specific heat measured under the applied fields ($\mu_0 H = 1$ to 9 T) from the specific heat data measured at $\mu_0 H = 0$ T.

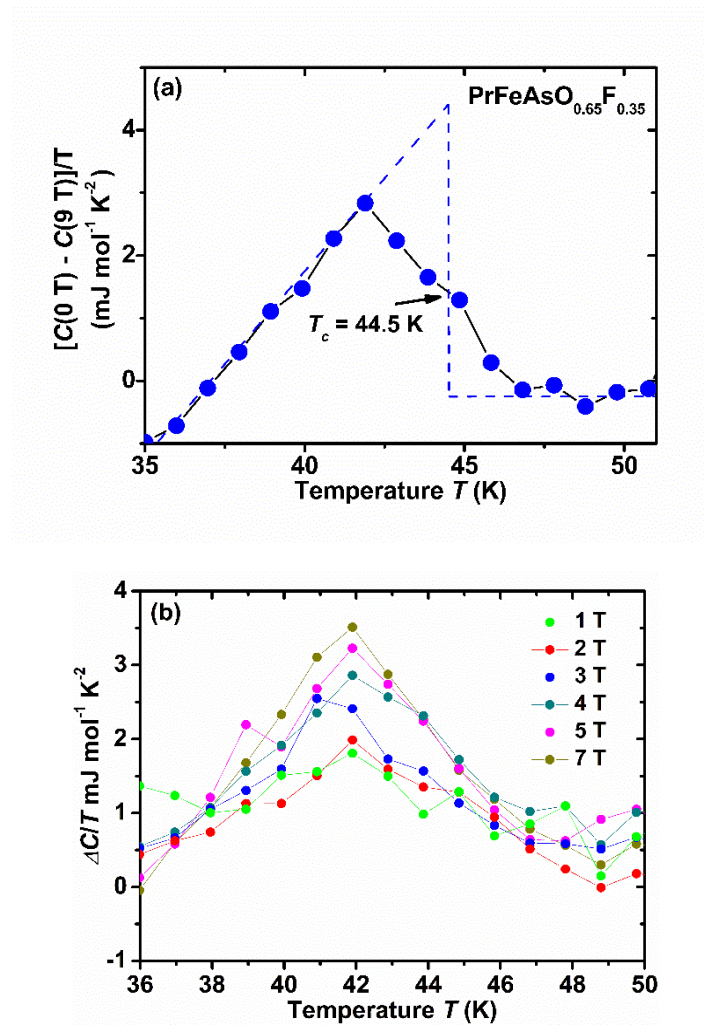

**Figure 9:** Electronic phase diagram of PrFeAsO$_{1-x}$F$_x$. The Neel temperature associated with Fe-spin-density-wave (SDW) order: $T_N^{Fe}$ (square symbols) and the magnetic ordering temperature of Pr, $T_N^{Pr}$ (triangle symbols), are determined from the resistivity measurements of the parent ($x = 0$) and lightly doped ($x = 0.1$) samples. The circle symbols represent the superconducting $T_c^{onset}$, extracted from the resistivity measurements for F-doped Pr1111 samples with $0.15 \leq x \leq 0.7$. The open diamond symbols correspond to the $T_c^{onset}$ obtained from the magnetization measurements for the samples $x = 0.2, 0.25, 0.3, 0.35, 0.4$ and $0.5$, while the inverted open triangle represents the superconducting transition determined from the specific heat data for the sample $x = 0.35$. Samples with $x = 0.8$ and $1.0$ exhibit no superconductivity and crystallize in a cubic phase, indicated by solid hexagon symbols.

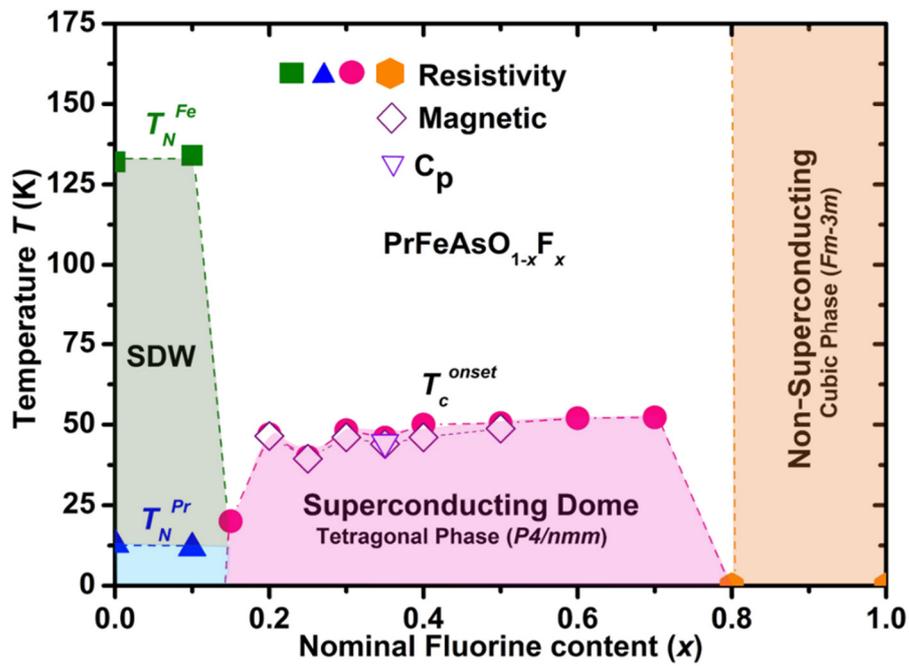

<u>Supplementary Material</u>

# Complete electronic phase diagram and enhanced superconductivity in fluorine-doped PrFeAsO$_{1-x}$F$_x$

Priya Singh[1], Konrad Kwatek[2], Tatiana Zajarniuk[3], Taras Palasyuk[2], Cezariusz Jastrzębski[2], A. Szewczyk[3], Shiv J. Singh[1]*

[1]*Institute of High Pressure Physics (IHPP), Polish Academy of Sciences, Sokołowska 29/37, 01-142 Warsaw, Poland*

[2]*Faculty of Physics, Warsaw University of Technology, Koszykowa 75, 00-662, Warsaw, Poland*

[3]*Institute of Physics, Polish Academy of Sciences, Aleja Lotnikow 32/46, 02-668 Warsaw, Poland*

**\*Corresponding author:**
 *Email:* sjs@unipress.waw.pl
 https://orcid.org/0000-0001-5769-1787

**Table T1:** Sample dimensions and distance between voltage contacts for PrFeAsO$_{1-x}$F$_x$ samples used in resistivity measurements, performed via the four-probe method as described in the Experimental section.

| Sample composition PrFeAsO$_{1-x}$F$_x$ | Distance between the voltage wires, $l$ (mm) | Thickness of the sample, $t$ (mm) | Width of the sample, $w$ (mm) |
|---|---|---|---|
| $x = 0$ | 2.12 | 2.60 | 3.63 |
| $x = 0.1$ | 2.15 | 1.34 | 4.26 |
| $x = 0.15$ | 1.89 | 2.37 | 3.19 |
| $x = 0.2$ | 1.38 | 2 | 2.01 |
| $x = 0.25$ | 3.2 | 2.12 | 3.54 |
| $x = 0.3$ | 3.74 | 1.41 | 4.94 |
| $x = 0.35$ | 2.6 | 2.52 | 4.41 |
| $x = 0.4$ | 1.92 | 3.03 | 3.09 |
| $x = 0.5$ | 2.61 | 2.72 | 3.61 |
| $x = 0.6$ | 1.69 | 1.99 | 2.05 |
| $x = 0.7$ | 1.99 | 1.98 | 3.11 |
| $x = 0.8$ | 1.99 | 1.96 | 3.09 |
| $x = 1$ | 1.98 | 2.04 | 3.25 |

# B1: Electrical Resistivity Analysis:

**Figure S1:** Temperature dependence of $d\rho/dT$ for PrFeAsO$_{1-x}$F$_x$ with $x = 0$ and $x = 0.1$. The higher-temperature anomaly ($T_s$) corresponds to the structural transition from tetragonal to orthorhombic symmetry, consistent with observations in other 1111 compounds [1], [2], [3], [4]. The lower-temperature kink indicates the antiferromagnetic transition of Fe ($T_N^{Fe}$).

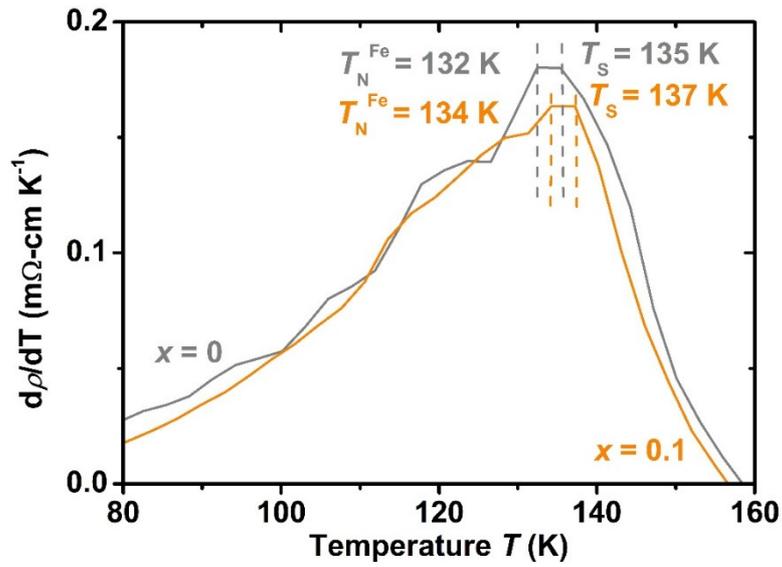

**Figure S2:** Variation of the superconducting onset transition temperature $T_c$ as a nominal fluorine content ($x$), comparing the present study with previous results reported by Rotundu *et al.* [5].

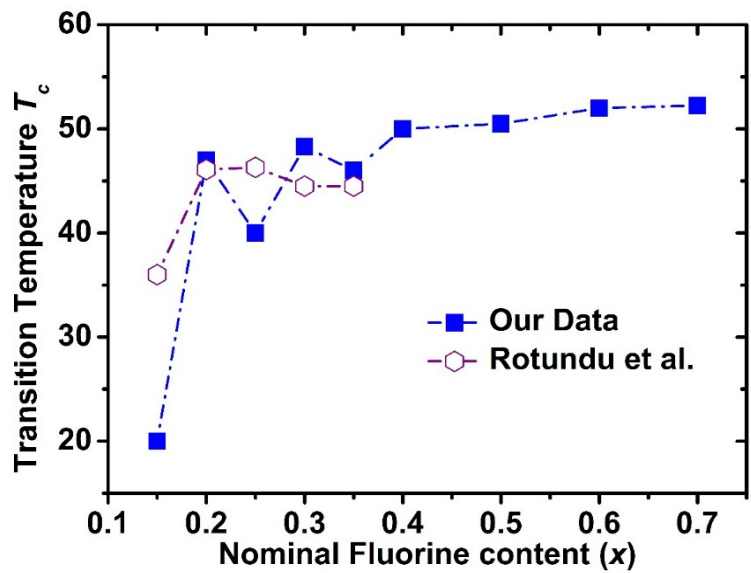

**Figure S3:** The temperature dependence of the resistivity ($\rho$) (logarithmic scale) for PrFeAsO$_{1-x}$F$_x$ samples in the low-temperature region (7-60 K). Panel (**a**) shows the compositions $x = 0.15$ to 0.4, and panel (**b**) represents the higher fluorine contents $x = 0.5$ to 1.0.

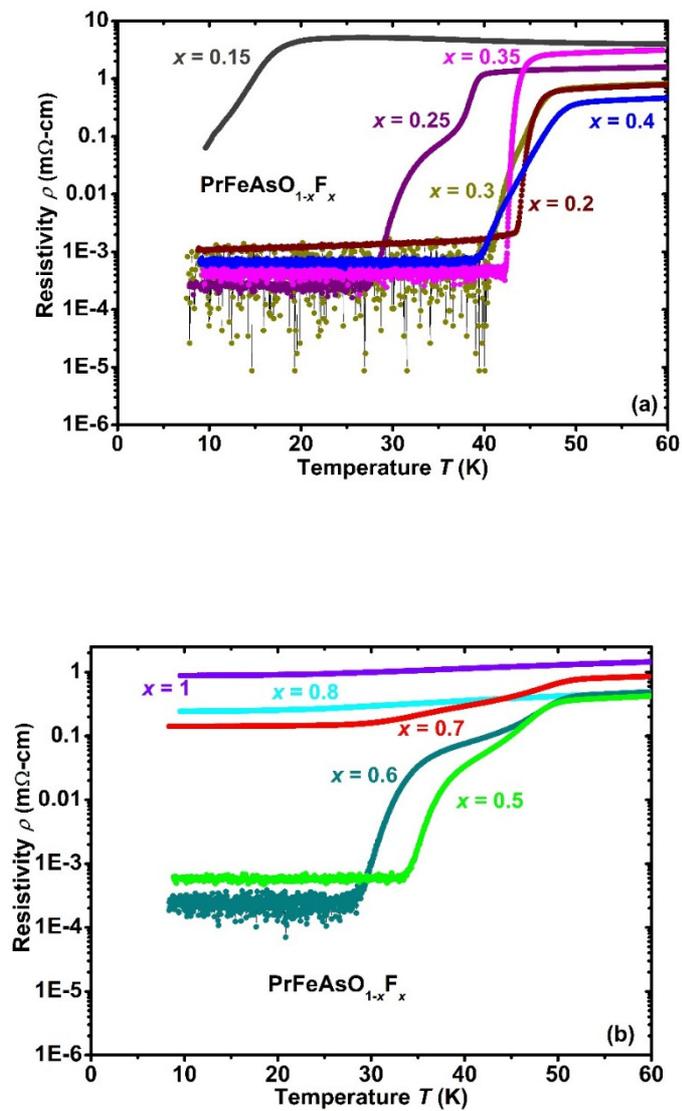

# B2: Normal-State Transport Fitting:

To further understand the normal-state transport behaviour, we have analyzed the zero-field resistivity data of the PrFeAsO$_{1-x}$F$_x$ samples in the temperature range of 55-300 K using the power-law relation $\rho = \rho_0 + AT^n$, where $\rho_0$ represents the residual resistivity at zero temperature and $A$ is the temperature coefficient. For a consistent comparison across different doping levels, the resistivity data were normalized by the room temperature resistivity value, i.e., $\rho_{300K}$, and fitted using the expression $\rho_{normalized} = AT^n$ for the samples $x = 0.2$ to 0.5. Since the entire temperature range could not be accurately described by a single fitting parameter, the data are analysed in four distinct temperature intervals: 55 to 80 K, 80 to 120 K, 120 to 220 K, and 220 to 300 K. A similar multiregional fitting approach has been previously applied to the FeSe(11) superconductor [6]. In Figures S4(a) and S4(b), we have shown the normalized resistivity in the low temperature range just above $T_c$: 55 K ≤ T ≤ 80 K with the variation of $T^{0.8}$ dependence ($n_1$ ~0.8) and in the high temperature range, 120 K ≤ T ≤ 220 K with the variation of $T^{1.3}$ dependence ($n_3$ ~1.3). The red dotted lines in these figures represent the resistivity fittings performed in the temperature ranges of 55 to 80 K and 120 to 220 K, respectively, and the corresponding extracted exponents, $n_1$ and $n_3$ for each sample with $x = 0.2$–0.5, are presented in Figure S5. Similarly, the coefficients $n_2$ and $n_4$ are obtained for the intermediate (80–120 K) and high-temperature (220–300 K) regions, respectively. Based on this analysis, we have plotted the variation of the resistivity coefficient $n$ across the whole temperature region with respect to fluorine doping contents ($x$) and the temperature range, displayed as a color map in Figure S4(c). It is evident that $n$ remains consistently below 2 ($n < 2$) throughout the entire temperature range and doping series, clearly indicating a deviation from conventional Fermi-liquid behavior in the PrFeAsO$_{1-x}$F$_x$ system. Notably, a $T^2$ dependence, typically associated with Fermi-liquid behavior, is not observed even near room temperature. This observation contrasts with previous reports of $T^2$ re-emergence in other iron-based superconductors such as FeSe$_{1-x}$S$_x$ [6], [7], [8], BaFe$_2$(As$_{1-x}$P$_x$)$_2$ [9], [10] and partially for LaFeAsO$_{1-x}$H$_x$ [1].

Furthermore, the extracted $A$ coefficient is designated as $A_1$, $A_2$, $A_3$, and $A_4$ corresponding to the temperature intervals of 55-80 K, 80-120 K, 120-220 K, and 220-300 K, respectively, which are associated with the resistivity exponents $n_1$, $n_2$, $n_3$, and $n_4$. The variations of $A_1$ and $A_3$ with fluorine content ($x$) are shown in Figures S4(d) and S4(e), while the corresponding $A_2$ and $A_4$ dependencies are presented in the Figure S6. A clear divergence in the exponent $n$ is observed around $x \approx 0.35$ across all temperature ranges, indicating the presence of an optimal doping region near this composition in the PrFeAsO$_{1-x}$F$_x$ system. The extracted *A* coefficient in the normal state (55–300 K) ranges from approximately 10 to 300 $\mu\Omega$-cm K$^{-2}$, which is significantly higher than the values reported for single-crystalline FeSe$_{1-x}$S$_x$ [7]. Notably, such a systematic correlation between the coefficient *A* and the fluorine doping content over a wide temperature range has not been previously established for the 1111 family of iron-based superconductors. Recently, Jiang et al. [11] reported a similar anomalous linear-temperature (*T*) dependence of resistivity over a wide range of doping in electron-doped FeSe, with $A_1 \propto T_c^2$, suggesting a link between the so-called "strange-metal" scattering and the superconducting pairing mechanism. This relationship mirrors behaviors observed in unconventional systems such as cuprates and Bechgaard salts, implying a broader, possibly universal, non-Fermi-liquid character among correlated superconductors. In the present study, the pronounced enhancement of $A_1$ and $A_3$ near *x* = 0.35 qualitatively reflects an increase in the effective electron mass $m_e^*$ as $A \propto (m_e^*)^2$ [10], [12], [13] according to the Kadowaki-Woods relation. This divergent behavior of *A* signifies a substantial enhancement in quasiparticle mass and density of states near the optimal doping level. Such non-Fermi-liquid behavior has been theoretically interpreted within the spin-fermion framework [14], where a limited portion of the Fermi surface, referred to as "hot spots," plays a dominant role in scattering and superconducting pairing. Alternatively, several studies have proposed that unequal participation of electron and hole bands near the Fermi energy may give rise to regions of enhanced scattering (hot spots) and reduced scattering (cold spots), collectively producing the observed non-Fermi-liquid characteristics [6]. Overall, the evolution of the *A* coefficient with fluorine content reinforces the three distinct doping regimes—underdoped, optimally doped, and overdoped—defined earlier from the superconducting transition behavior (Figure 4(b)–(d)). The distinct divergence of both $T_c$ and the coefficient *A* near *x* ≈ 0.35 strongly corroborates the boundary of the optimally doped region in PrFeAsO$_{1-x}$F$_x$.

Following this, resistivity fittings are also performed for the higher fluorine-doped samples within the range 0.5 < *x* < 0.8. The analysis revealed resistivity exponents *n* < 1 and *A* coefficients on the order of 100 $\mu\Omega$-cm K$^{-2}$, indicating pronounced deviations from Fermi-liquid behavior. Such sublinear temperature dependence suggests the presence of strong electronic fluctuations, likely arising from excessive fluorine incorporation. Moreover, the enhanced effective electron mass $m_e^*$ inferred from the large *A* values implies a substantial increase in the density of states near the Fermi energy, $\varepsilon_F$. This enhancement can strengthen Cooper pairing interactions, thereby contributing to the observed rise in the superconducting transition temperature $T_c$ from 48 K at $x = 0.30$ to approximately 52.3 K for higher fluorine concentrations. However, the origin of this non-Fermi-liquid behavior, characterized by a nearly linear $T$-dependence of resistivity, is widely attributed to low-energy electronic fluctuations mediated by spin-fermion interactions rather than conventional phonon coupling [14]. The strengthening of these spin-fermion interactions enhances pairing but, when excessively strong, may reduce coherence and ultimately suppress $T_c$ or destabilize the superconducting phase. Consistent with this picture, $T_c$ saturates around 51–52 K for $0.5 < x \leq 0.7$, indicating a solubility limit of fluorine that constrains further improvement in superconducting properties. This saturation behavior aligns with previous findings [14], which propose that $T_c$ becomes limited by a crossover from hot-spot or spin-fermion-dominated pairing to pairing governed by the overall Fermi surface topology. For samples with $x > 0.7$, superconductivity completely disappears, suggesting the emergence of a symmetric Fermi surface associated with the cubic *Fm-3m* structure observed in the XRD analysis (Figure 1). This structural transition likely arises from strong lattice disorder and local inhomogeneity caused by heavy fluorine substitution at oxygen sites, leading to the suppression of superconductivity and stabilization of a non-superconducting cubic phase.

**Figure S4:** The plot of normalized resistivity $\rho/\rho_{300\,K}$ as a function of the temperature **(a)** $T^{0.8}$ in the range 55 K ≤ T ≤ 80 K and **(b)** $T^{1.3}$ in the range 120 K ≤ T ≤ 220 K. The used red lines represent linear fits and serve as a guide to the eye. Additional fittings for the coefficient $n_1$ and $n_3$ for these temperature regions are provided in the Figure S5. **(c)** Color map depicting the variation of resistivity exponent $n$, obtained from the power-law relation $\rho_{normalized} = AT^n$ across the temperature ranges: 55 K ≤ T ≤ 80 K, 80 K ≤ T ≤ 120 K, 120 K ≤ T ≤ 220 K and 220 K ≤ T ≤ 300 K. Panels **(d)** and **(e)** show the variation of coefficients $A_1$ and $A_3$, extracted from the power-law $\rho_{normalized} = AT^n$ fitting in the temperature ranges 55 K ≤ T ≤ 80 K and 120 K ≤ T ≤ 220 K, respectively. The variation of coefficients $A_2$ and $A_4$ is provided in the Figure S6.

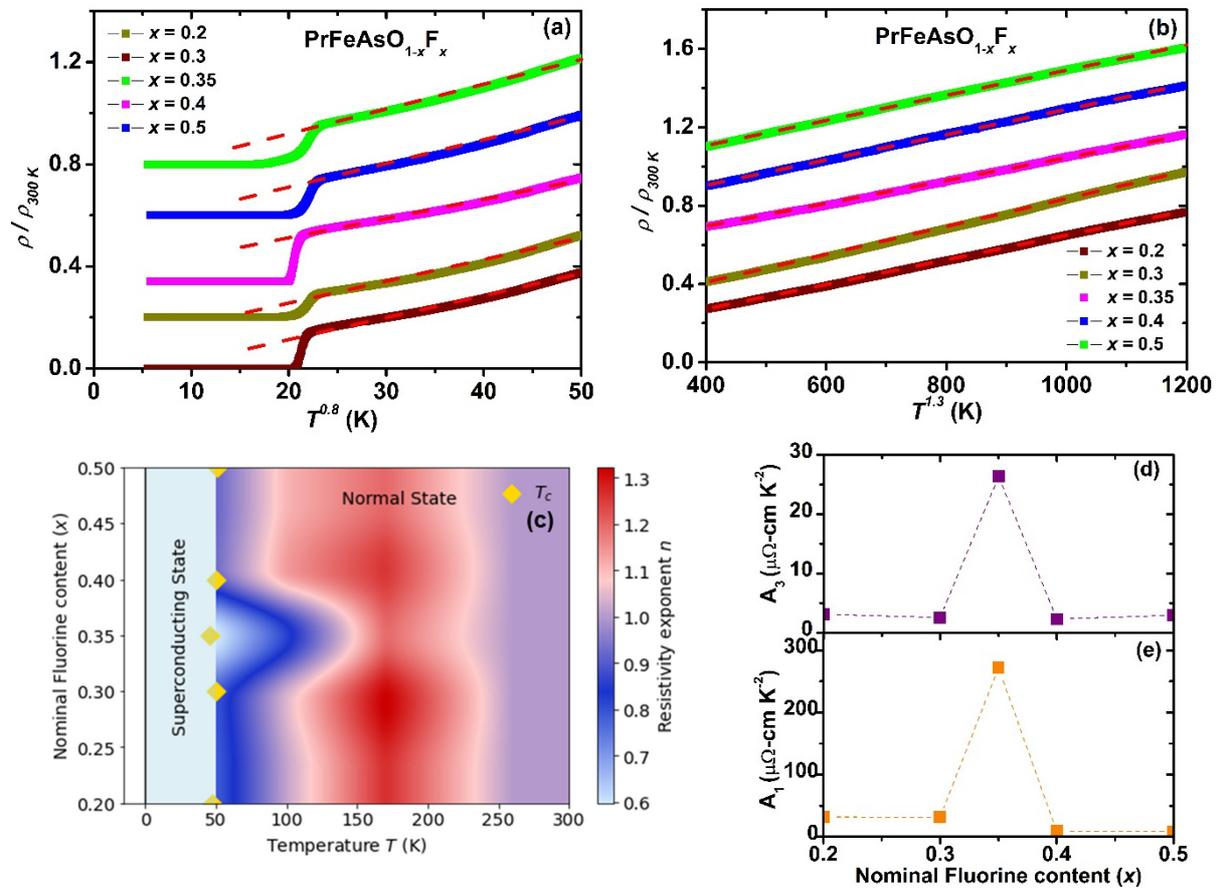

**Figure S5: (a)–(e)** Power-law fitting of the normal state resistivity $\rho = AT^n$ for PrFeAsO$_{1-x}$F$_x$ samples, where $n_1$ and $n_3$ are resistivity exponents extracted in the temperature ranges 55–80 K and 120–220 K, respectively.

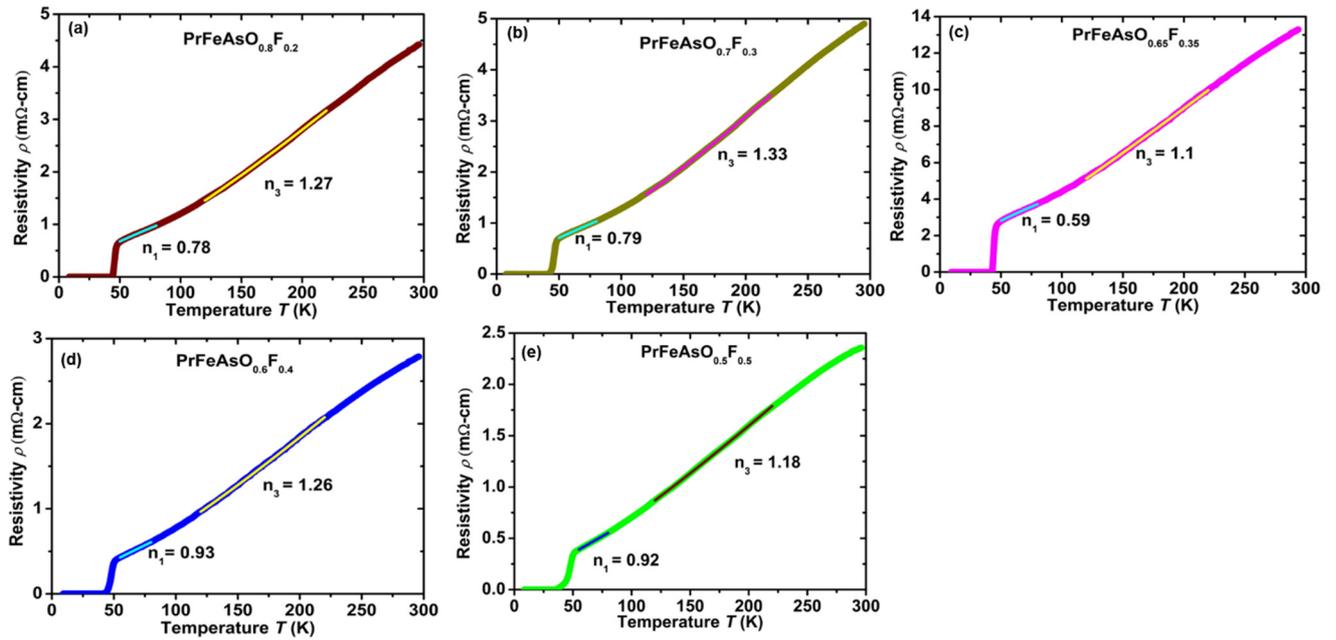

**Figure S6:** The variation of the coefficients **(a)** $A_2$ and **(b)** $A_4$ obtained from the Power law fitting $\rho = AT^n$ for PrFeAsO$_{1-x}$F$_x$ samples with respect to the nominal fluorine content $x$. The coefficients correspond to the resistivity exponents $n_2$ and $n_4$, and are extracted from the temperature ranges 80–120 K and 220–300 K, respectively, as a function of the nominal fluorine content.

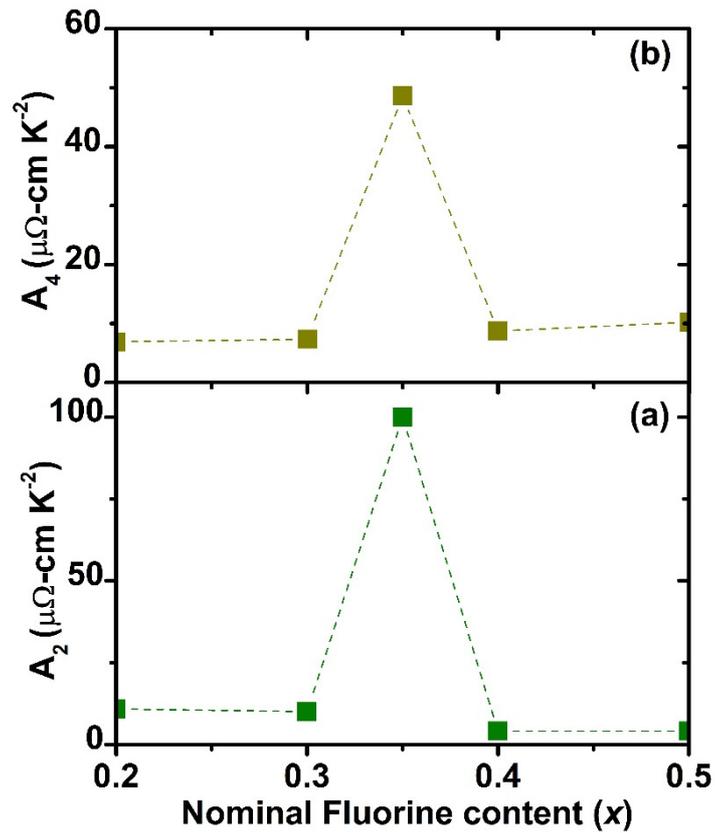

# B3: Magnetization Susceptibility and Critical Current Density ($J_c$):

**Figure S7: (a)** Temperature dependence of the normalized magnetic moment $M/M_{5K}$ measured under zero-field cooling (ZFC) and field cooling (FC) conditions. The arrow indicates the procedure used to determine the superconducting onset transition temperature from the magnetization data. **(b)** The variation of critical current density $J_c$ as a function of the applied magnetic field $\mu_0H$ at $T = 5$ K for the samples $x = 0.2, 0.25, 0.3, 0.35, 0.4$ and $0.5$. **(c)** The dependence of critical current density $J_c$ on the applied magnetic field $\mu_0H$ at $T = 5$ K for two independent pieces of PrFeAsO$_{0.8}$F$_{0.2}$ and PrFeAsO$_{0.7}$F$_{0.3}$ samples prepared from the same batch, demonstrating reproducibility of the measurements.

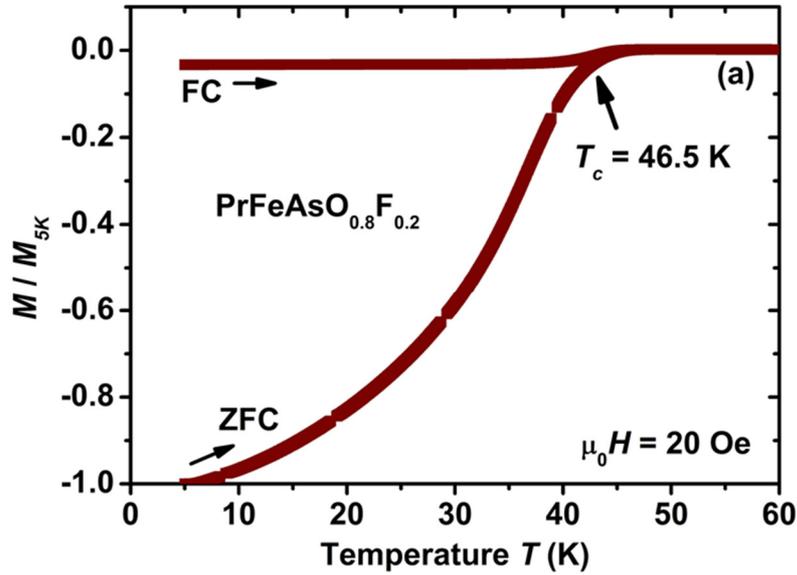

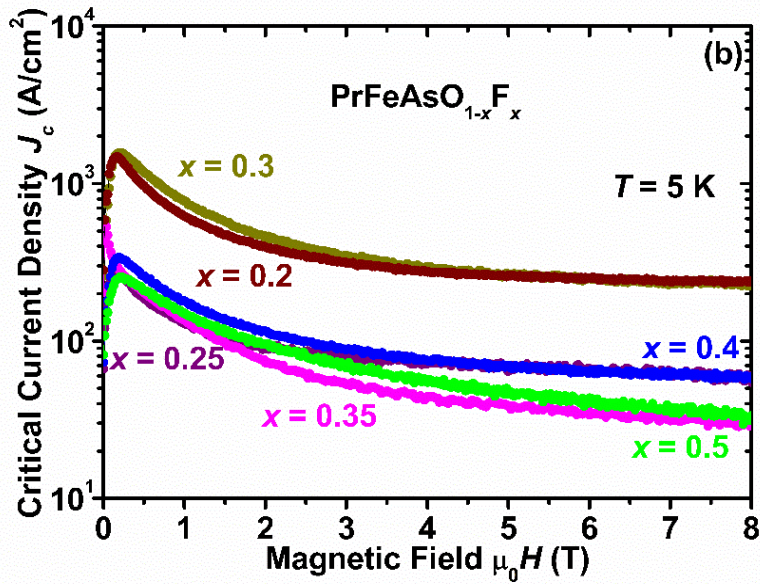

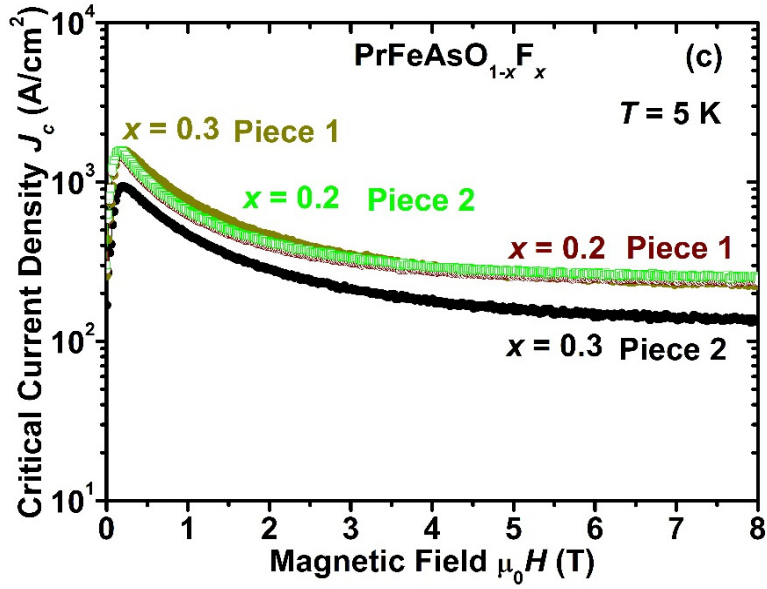

The critical current density $J_c$ is a key parameter determining the performance of a superconductor, especially for potential technological applications. We have measured the magnetic hysteresis loop at 5 K for the PrFeAsO$_{1-x}$F$_x$ samples with $x$ = 0.2, 0.25, 0.3, 0.35, 0.4 and 0.5 to evaluate their current-carrying capabilities. The width of the magnetic hysteresis loop ($\Delta m$) was determined by subtracting the magnetization values recorded during the increasing field from those during the decreasing field. The $J_c$ values were then calculated using the Bean critical state model: $J_c = 20 \frac{\Delta m}{Va\left(1-\frac{a}{3b}\right)}$, where $a$ and $b$ represent the sample dimensions ($a < b$), and $V$ denotes the sample volume [15]. The calculated $\Delta m$ values and corresponding dimensions were used to estimate $J_c$ for the selected PrFeAsO$_{1-x}$F$_x$ compositions, and the results are plotted in Figure S7(b) and Figure S7(c). Among the studied samples, the $x$ = 0.20 and 0.30 compositions exhibit the highest $J_c$ values (~ 1 × 10$^3$ A/cm$^2$) in the low magnetic fields, followed by a gradual decrease with increasing field strength. Further increasing fluorine doping levels i.e., the samples $x$ = 0.25, 0.35, 0.4 and 0.5 display a similar field dependence but slightly lower $J_c$ values compared to the samples $x$ = 0.2 and 0.3, indicating a reduction in flux-pinning effectiveness. In particular, the highly fluorine-doped sample ($x$ = 0.50) displays a more rapid suppression of $J_c$ at higher magnetic fields, likely due to the increased presence of secondary phases, as noted earlier in the XRD analysis. Overall, all samples demonstrate relatively high $J_c$ values in low magnetic fields, followed by a moderate decrease with increasing magnetic field strength. To confirm the reproducibility, magnetic hysteresis measurements were repeated for two separate pieces of some selected compositions. The independent pieces exhibited nearly identical $J_c$ values and field-dependent behaviour, as shown in Figure S7(c). The slightly enhanced $J_c$ values observed for $x = 0.20$ and 0.30 may be attributed to improved grain connectivity in these samples, which promotes stronger flux pinning compared to the other samples. The magnitude and field dependence of the $J_c$ values obtained in this study are consistent with those reported for bulk iron-based superconductors, further confirming the intrinsic nature of the observed behaviour [16]. Interestingly, the $J_c$ behaviour of these F-doped PrFeAsO$_{1-x}$F$_x$ samples exhibits a relatively weak dependence on the applied magnetic field, similar to that reported for single crystal PrFeAsO$_{0.60}$F$_{0.35}$ [17], suggesting their potential suitability for the magnetic applications. However, the $J_c$ values obtained for these polycrystalline PrFeAsO$_{1-x}$F$_x$ samples are approximately two to three orders of magnitude lower than those reported values (~$10^5$ A/cm$^2$) for the single-crystal PrFeAsO$_{0.60}$F$_{0.35}$ [17], indicating that further improvements in bulk sample quality and microstructural homogeneity are necessary to enhance their current-carrying performance.

## B4: Pinning Force ($F_p$) Analysis:

Generally, the critical current density is closely related to the vortex pinning force ($F_p$), which can be expressed as $F_p = \mu_0 H \times J_c$, where $\mu_0$ is the vacuum permeability. Based on the $J_c$ values shown in Figure S7(b), the maximum pinning force $F_p$ reaches approximately $10^7$ N/m$^3$ for the $x = 0.3$ and 0.2 samples, which is about one order of magnitude higher than that of the $x = 0.25$ and 0.4 samples, as shown in the Figure S8. This enhanced pinning force might be a reason for the high $J_c$ value observed in the samples $x = 0.3$ and 0.2 compared to other samples. These $F_p$ values are comparable to those reported for other members of the 1111 family [16], [18]. As shown in the Figure S8, the pronounced pinning force $F_p$ values at higher magnetic fields suggests a dominant contribution from intragranular pinning mechanisms, similar to that observed for polycrystalline SmFeAsO$_{0.8}$F$_{0.2}$ [18] and PrFeAsO$_{0.6}$F$_{0.12}$ [16]. The reported scaling analysis of the pinning force $F_p(H)$ indicates an enhancement of $\delta T_c/\delta l$-type pinning at high magnetic fields, potentially arising from dopant inhomogeneities in the F/O distribution, spatial variations in the mean free path ($l$) near defects, point defects, or nanoscale disorder commonly observed in the 1111 family [19].

**Figure S8:** The magnetic field dependence ($\mu_0 H$ = 0-9 T) of the pinning force $F_p$ is illustrated for PrFeAsO$_{1-x}$F$_x$ with $x$ = 0.2, 0.25, 0.3 and 0.4. For the composition $x$ = 0.3 and 0.2, data is plotted from piece 1 (see figure S7(c)).

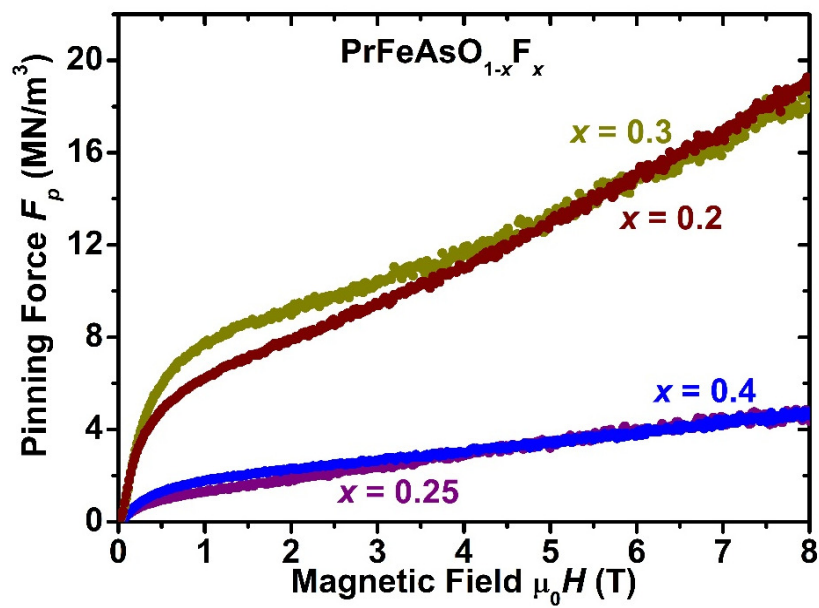